\documentstyle[epsf,cite]{areviews}

\begin{document}

\pagestyle{myheadings} \markboth{\rm JENKINS}{\rm LARGE-$N_c$ BARYONS}
\title{LARGE-$N_c$ BARYONS}
\author{Elizabeth Jenkins
\affiliation{Department of Physics, 
University of California at San Diego, La Jolla, California 92093, 
ejenkins@ucsd.edu}}


\begin{abstract}
A spin-flavor symmetry emerges for baryons 
in the large-$N_c$ limit.  
Large-$N_c$ baryons form irreducible representations of the 
spin-flavor algebra, and their static properties can be computed in a 
systematic expansion in $1/N_c$.  Symmetry relations for static baryon
matrix elements are obtained
at various orders in the $1/N_c$ expansion
by neglecting subleading $1/N_c$ corrections.
Equivalent relations arise in the quark and Skyrme models, which
satisfy the same large-$N_c$ group theory as QCD.  
The $1/N_c$ expansion yields useful results for QCD baryons with $N_c=3$. 
\end{abstract}

\maketitle

\def\clebsch#1#2#3#4#5#6{\left(\matrix{#1&#3\cr#2&#4\cr}\right.\left|
\matrix{#5\cr#6\cr}\right)}
\def\sixj#1#2#3#4#5#6{\left\{\matrix{#1&#2&#3\cr#4&#5&#6\cr}\right\}}
\def\openone{\leavevmode\hbox{\small1\kern-3.3pt\normalsize1}}
\def\slash#1{\rlap{$#1$}/} 

\section{INTRODUCTION}

Quantum chromodynamics, the theory of the strong interactions, is
an $SU(3)$ gauge theory of quarks and gluons.  
At low energies, the running coupling constant of the theory is large, 
and the colored quarks and gluons are confined into colorless
mesons and baryons.
It is not known how to calculate the structure and interactions of mesons and
baryons directly in terms of the underlying quark-gluon dynamics because the
theory is strongly coupled at low energies.
The large $N_c$ formulation of QCD~\cite{thooft}
provides a framework for studying 
the nonperturbative QCD dynamics of hadrons
in a systematic expansion in the parameter $1/N_c$, where $N_c$ is the number
of colors.    
Recently, significant progress has been made in the study of baryons in
$1/N_c$. 
A spin-flavor symmetry arises for large-$N_c$ baryons,
and can be used to classify large-$N_c$ baryon states and matrix
elements.
The spin-flavor structure of the baryon $1/N_c$ expansion is calculable,
and yields predictions for the static properties of baryons for any
$N_c$.
These new developments will be the focus of this review.
We begin with a brief introduction of large $N_c$ QCD before turning to this
work.  For surveys of large $N_c$ QCD, see 
Refs.~\cite{witten,coleman,leshouches}.

Large $N_c$ QCD is the $SU(N_c)$ gauge theory of quarks and gluons, where
the number of colors $N_c$ is a parameter of the theory~\cite{thooft}.  
For large $N_c$, the number of gluons is $N_c^2 -1 =O(N_c^2)$,
whereas the number of quarks is $O(N_c)$, so gluons are much more prevalent
than quarks. 
An analysis of the $N_c$-dependence of quark-gluon diagrams is required to
define the large-$N_c$ limit of the theory.
The leading $N_c$-dependence of quark-gluon diagrams can be determined by
replacing the gluon gauge field $(A^{\mu})^A$ in the adjoint color 
representation by a tensor $(A^{\mu})^i_j$ with one lower index and 
one upper index in the fundamental color representation, as depicted in
Fig.~1.
When Feynman diagrams are drawn using double line notation, the power of $N_c$
of a diagram is obtained by counting the number of closed quark loops
since each quark loop implies an unconstrained summation 
over the color index of the quark, which produces a factor of $N_c$.
't Hooft analyzed the $g$- and $N_c$-counting of vacuum Feynman diagrams 
in double line notation, and
found that the graphs are proportional to
\begin{equation}      
\left(g^2 N_c\right)^{\frac 1 2 V_3 + V_4} N_c^{\chi},
\end{equation} 
where $g$ is the coupling constant of the theory, $V_n$ are the number of
$n$-point vertices of quarks and gluons in a given diagram, and 
the Euler character $\chi = 2 -2H -L $
is a topological invariant
depending of the number of holes $H$ and the number of quark loops $L$
of the diagram.  
Diagrams with large numbers of vertices grow with arbitrarily large
powers of $N_c$ unless the large-$N_c$ limit $N_c \rightarrow \infty$ is taken with $g^2 N_c$
held fixed.
This constraint can be implemented by performing the rescaling
\begin{equation}\label{gscaling}
g \rightarrow {g \over \sqrt{N_c} }
\end{equation}
in the large-$N_c$ QCD Lagrangian.  
After this rescaling, all $N_c$-dependence of the
theory is manifest, and the new coupling constant $g$ is order unity.  
The running of the coupling constant is
governed by the $\beta$ function of large-$N_c$ QCD.  All $N_c$-dependence
factors out of the large-$N_c$ QCD $\beta$ function equation under the
rescaling Eq.~(\ref{gscaling}), so the rescaled coupling constant
$g$ of the large-$N_c$ theory becomes strong at the same scale
$\Lambda_{\rm QCD}$ as the $N_c=3$ theory.
For energies $E \le O(\Lambda_{\rm QCD})$, the large-$N_c$ theory is
strongly coupled, and is presumed to exhibit confinement.   

Most of the initial work on large-$N_c$ QCD concentrated on
mesons~\cite{thooft,ven1,ven2}.  
Large-$N_c$ mesons are color singlet bound states of a quark and an
antiquark,
\begin{equation}\label{meson} 
\sum_{i=1}^{N_c} \bar q_i q^i,
\end{equation}
where the color summation is given explicitly.
The $N_c$-dependence of meson amplitudes can be determined by studying
quark-gluon diagrams.
The leading order graphs have a single quark loop and
no holes, so they can be written in a plane in double line notation.
Arbitrary numbers of planar gluons can be exchanged inside the single quark
loop without affecting the $N_c$-counting of the diagram.
Diagrams with additional quark loops or nonplanar gluons are systematically
suppressed in $1/N_c$.  
Each additional quark loop is suppressed by $1/N_c$ and each
nonplanar gluon is suppressed by $1/N_c^2$. 
Meson self-couplings are analyzed by studying planar diagrams with
meson insertions on the quark loop (see Fig.~2).  
Each meson insertion is accompanied by a
factor of $1/\sqrt{N_c}$, which is the normalization factor for the meson
quark operator Eq.~(\ref{meson}).  The coupling of $n$-mesons is given by
the planar diagrams with $n$ meson insertions on the quark loop.  There is a
single factor of $N_c$ from the single quark loop of the diagram and $n$
factors of $1/\sqrt{N_c}$, so the $n$-meson coupling is $O(N_c^{1 - n/2})$.
From this $N_c$ counting, one concludes that a meson decay constant is
$O(\sqrt{N_c})$; the meson mass is $O(1)$; 
the self-coupling of three mesons is $O(1/\sqrt{N_c})$; the self-coupling
of four mesons
is $O(1/N_c)$; etc.  The amplitude for a meson to decay to two other
mesons is $O(1/\sqrt{N_c})$, so the decay width is $1/N_c$.  Thus,
large-$N_c$
mesons are narrow states which are weakly coupled to each other.  
In the large-$N_c$ limit, mesons are stable and non-interacting.
        
Baryons in large-$N_c$ QCD were studied by
Witten~\cite{witten}.  Large-$N_c$ baryons are color singlet bound states
of $N_c$ valence quarks.  The number of quarks in a baryon grows with 
$N_c$, and changes as the parameter $N_c$ is varied.    
The color indices of the $N_c$ quarks
are contracted with the $SU(N_c)$ color $\epsilon$-symbol to form a color
singlet state,
\begin{equation}
\epsilon_{i_1 i_2 i_3\cdots i_{N_c}}\ q^{i_1} q^{i_2} q^{i_3} 
\cdots q^{i_{N_c}},
\end{equation}
so a large-$N_c$ baryon is totally antisymmetric in the color indices 
of its $N_c$ valence quarks.
At leading order, the large-$N_c$ baryon is described
by quark-gluon diagrams consisting of $N_c$ valence quark lines with arbitrary
planar gluons exchanged between the quark lines.
Each sea-quark loop is suppressed by $1/N_c$ and each nonplanar
gluon exchange is suppressed by $1/N_c^2$.  
(For a description of planar baryon diagrams, see Ref.~\cite{leshouches}.
Ironically, planar diagrams for baryons cannot always be drawn in a plane.)
The baryon mass grows with $N_c$,
\begin{equation}
M_{\rm baryon} \sim O(N_c) \ ,
\end{equation}
and baryons become infinitely heavy in the large-$N_c$ limit.     
For massless quarks, the only dimensional parameter of the 
large-$N_c$ QCD theory is $\Lambda_{\rm QCD}$, and
$M_{\rm baryon} \sim N_c \Lambda_{\rm QCD}$. 

The number of quarks inside a large-$N_c$ baryon grows as
$N_c$ in the large-$N_c$ limit, 
but the size of the baryon is governed by $\Lambda_{\rm QCD}$ and 
remains fixed.  Thus, as $N_c$ becomes large, the quarks in the baryon become
denser and denser.  Witten realized that this situation could be described by a
Hartree approximation in which each quark in the $N_c$-quark bound state
interacts with the average potential produced by the other $O(N_c)$ 
quarks~\cite{witten}.  In the large-$N_c$ limit, the Hartree
picture becomes exact, and each quark experiences 
the same average Hartree potential. 
When $N_c$ is very large, the Hartree Hamiltonian for a large-$N_c$ baryon
is given by the sum of $N_c$ identical quark Hamiltonians
\begin{equation}
{\cal H}_{\rm baryon}= N_c {\cal H}_{\rm quark}\ ,
\end{equation}
where ${\cal H}_{\rm quark}$ is the Hamiltonian for any one quark in the
baryon, and is $O(1)$.  The $N_c$-dependence of the Hartree Hamiltonian
is trivial for large $N_c$, so instead of
solving for the baryon many-body wavefunction, one can consider the much
simpler problem of finding the wavefunction of a single quark.  The baryon
wavefunction is given by the product of $N_c$ identical quark wavefunctions, 
\begin{equation}
\Psi\left(x_1,\ldots,x_{N_c}\right)
=\prod_{i=1}^{N_c} \phi(x_i)  \ .
\end{equation}
For the lowest-lying ground state baryons, all of the quarks are in the same
rotationally invariant ground state solution of the Hartree Hamiltonian ${\cal
H}_{\rm quark}$, and the $N_c$ quark wavefunctions are combined in an $s$-wave, 
so the baryon wavefunction reduces to
\begin{equation}\label{hartree}
\prod_{i=1}^{N_c} \phi(r) \ .
\end{equation}
Note that the baryon wavefunction is 
totally antisymmetric in the $N_c$ color indices of the quarks, so it is
totally symmetric in the bosonic wavefunctions $\phi(r)$ of the quarks. 
The Hartree Hamiltonian can be given in closed form only for heavy quarks.  
However, one
expects the qualitative picture and the large-$N_c$ power counting to
remain valid for light quarks where the relativistic Hartree equations are
not known.

A second picture of large-$N_c$ baryons is obtained by considering the
weakly-coupled meson sector of large-$N_c$ QCD.  It has been speculated
that large-$N_c$ baryons 
are solitons~\cite{witten} of the large-$N_c$ meson Lagrangian
\begin{equation}
{\cal L}_{\rm meson} = N_c {\cal L}\left( \phi/\sqrt{N_c}\right) \ , 
\end{equation} 
where ${\cal L}$ is an arbitrary polynomial of the meson fields $\phi$.
The Skyrme model~\cite{skyrme,soliton,anw} is an explicit model realization of 
this idea 
using the chiral Lagrangian for pseudoscalar pions.
The soliton mass is order $N_c$, but its size and shape is independent of
$N_c$.  The large-$N_c$ couplings of mesons to baryons can be obtained by
expanding the action in the meson functional integral 
\begin{equation}
Z = \int {\cal D} \phi \ e^{i N_c S/ \hbar}, \qquad
S= \int d^4x \ {\cal L} \left( \phi/\sqrt{N_c}\right)
\end{equation}
about the baryon soliton configuration, and solving the
linear equations of motion
for the semiclassical meson field in the presence of a background 
baryon soliton~\cite{mattis}.

The $N_c$-dependence of baryon-meson scattering
amplitudes can be determined by studying quark-gluon diagrams.  
Typical graphs which contribute to the process 
baryon + meson $\rightarrow$ baryon + meson are given in Fig.~3,
where it is to be understood that
each of the $N_c$ quarks in the baryon has a different color from any
of the other quarks.  In Fig.~3(a), 
both mesons are inserted on the same quark line of the baryon,
so the color structure of the quark lines is preserved.
There are $O(N_c)$ possible choices for the quark line with the meson
insertions, and each meson couples to a   
quark-antiquark pair with amplitude $1/\sqrt{N_c}$, so the overall diagram is
$O(1)$.  Alternatively, the two mesons can be inserted on two different quark
lines of the baryon, as shown in Fig.~3(b).
In this case, a gluon must be exchanged between the two quark lines
to transfer energy between the incoming and outgoing mesons.  
The gluon interchanges the colors of quarks, so 
color conservation requires the exchange of the two quark lines.
There are $N_c (N_c -1)/2 = O(N_c^2)$ possible choices for the pair
of quark lines with meson insertions; each meson insertion is
accompanied by a factor of $1/\sqrt{N_c}$; and 
the gluon exchange is proportional to $g^2/N_c$, so the overall diagram 
is $O(1)$.
The generalization to multimeson-baryon scattering amplitudes
is straightforward.
In general, the
scattering amplitude for the process
baryon + meson $\rightarrow$ baryon + $(n-1)$ mesons is
$O(N_c^{1-n/2})$,
since each additional meson in the scattering process
reduces the scattering amplitude by a factor of
$1/\sqrt{N_c}$.

Mesons are strongly coupled to baryons for large $N_c$. 
Again, quark-gluon diagrams can be used to determine the
$N_c$-dependence of baryon-meson vertices.  
Typical graphs depicting the coupling of a
baryon to a single meson
are given in Fig.~4. 
The meson
can be inserted on a single quark line in the baryon
without affecting the color structure of the quark 
lines (Fig.~4(a)).  There are $O(N_c)$ possible
choices for the quark line with the meson insertion, and the meson insertion is
accompanied by a factor of $1/\sqrt{N_c}$, so the overall diagram is
$O(\sqrt{N_c})$.  Instead, the quark line of the meson can be reinserted
into the baryon at another position (Fig.~4(b)).  
This color rearrangement is mediated by
the exchange of a gluon.  There are $O(N_c^2)$ possible choices for the pair of
quark lines involved; the meson insertion gives a factor of $1/\sqrt{N_c}$; and
the gluon exchange yields a factor of $g^2/N_c$, so the overall diagram is
$O(\sqrt{N_c})$.
The large-$N_c$ power counting of vertices
containing a single baryon and $n$ mesons is determined from the same
quark-gluon diagrams analyzed for baryon-meson scattering amplitudes.
A baryon coupling to $n$ mesons is $O(N_c^{1-n/2})$,
since each additional meson in a vertex is accompanied by a factor 
of $1/\sqrt{N_c}$. 
 
(Note that the
above large-$N_c$ power counting rules apply for a single quark flavor.  
For more quark flavors, one must 
keep track of the flavor of each quark line in the quark-gluon diagrams. 
The flavor content of the mesons and baryons places restrictions on which quark
lines can be chosen for the meson insertions.
If the baryon contains $O(N_c)$ quarks of the relevant quark flavors, then
the $N_c$ counting remains the same.  However, if only $O(1)$ quarks in the
baryon are a relevant flavor, then further suppressions occur.)

Recently, it has been realized that the spin and flavor structure of
baryons simplifies for large $N_c$.  Consistency of the $N_c$-power counting
rules for baryon-meson scattering amplitudes and baryon-meson vertices leads
to nontrivial constraints on static
baryon matrix elements~\cite{dm,j}.  
Large-$N_c$ consistency conditions for the scattering of low-energy pions
with baryons imply that baryons satisfy a contracted spin-flavor algebra
in the large $N_c$ limit, a result derived by Dashen $\&$ Manohar~\cite{dm}
and by Gervais $\&$ Sakita~\cite{gs}.
Dashen $\&$ Manohar and Jenkins showed that subleading 
$1/N_c$ corrections are constrained by additional large-$N_c$ 
consistency conditions~\cite{dm,j}, so the   
spin-flavor structure of large-$N_c$
baryons with finite $N_c$ can be analyzed
in a systematic expansion in powers of $1/N_c$ about the large-$N_c$ limit.  
The spin-flavor structure and the order in $1/N_c$
of the subleading $1/N_c$ corrections are specified by the
algebraic constraints.  The subleading $1/N_c$ corrections break 
the leading order spin-flavor symmetry of large-$N_c$ baryons, and  
are parametrized by
operator products of the baryon spin-flavor generators~\cite{djm1}, which have
known baryon matrix elements.  The
baryon spin-flavor generators can be written 
as Skyrme model operators or nonrelativistic quark model 
operators~\cite{dm,djm1,djm2}.  
The formulation of the $1/N_c$ expansion 
using static quark operators has a natural connection to 
the Hartree picture~\cite{cgo} and to quark-gluon diagrams~\cite{lm}, while
the formulation using Skyrme model operators has a natural connection 
to the soliton picture and Skyrme model.
Symmetry relations amongst baryon couplings and
amplitudes are obtained by neglecting subleading terms in the $1/N_c$
expansion~\cite{dm,j,djm1,l,lmw,jm,djm2,jl,ddjm}.  
The large-$N_c$ 
Skyrme model~\cite{anw,guadagnini,bardakci,bijnens,manohar84,adkinsnappi,karlii,
mattis,kaplanklebanov} 
and the large-$N_c$ quark model~\cite{oldsu6,karli} 
both share the same spin-flavor group theory for baryons 
as large-$N_c$ QCD~\cite{gs,manohar84}, and yield equivalent 
model-independent predictions for baryons in an
expansion in $1/N_c$.  The recent work on the baryon $1/N_c$ expansion
distinguishes between group theoretic relations of the Skyrme and quark models
which are
model-independent and true in large-$N_c$ QCD, and 
results which are model-dependent and not consequences of large-$N_c$ QCD.
Connections to other large-$N_c$ models of baryons are found as
well~\cite{mattissilbar,doreyii,manohar94,doreymattis}.
It has been shown that       
the
baryon $1/N_c$ expansion can be usefully 
applied to QCD baryons with $N_c=3$. 
The complete pattern of spin-flavor breaking for QCD
baryons requires the inclusion
of $SU(3)$ flavor symmetry breaking, which is order $30\%$ and comparable to the
suppression factor $1/N_c$. 
A combined expansion in $1/N_c$ and $SU(3)$
flavor symmetry breaking has been performed for many baryon
quantities, including baryon axial 
couplings~\cite{dm,j,djm1,l,lmw,jm,djm2,jl,ddjm,jchpt},
masses~\cite{djm1,l,djm2,jl,jchpt,bedaque}, and magnetic 
moments~\cite{dm,broniowski,lmw,jm,ddjm}.
Other phenomenological applications of the baryon $1/N_c$ expansion include
heavy quark baryons containing a single charm or bottom quark~\cite{j,jhqet},
baryon chiral perturbation
theory~\cite{dm,j,djm1,lm,lmw,manohar94,jchpt,bedaque},
excited baryons~\cite{cgkm,py,py2},
and nuclear physics~\cite{kaplansavage,kaplanmanohar}.  Additional related work
can be found in Refs.~\cite{takamura,collinsgeorgi,lam}.

\section{SPIN-FLAVOR SYMMETRY FOR BARYONS}
\subsection{Large-$N_c$ Consistency Conditions}

Large-$N_c$ constraints on baryon static matrix elements are derived by
considering baryon-meson scattering at low energies $E \sim O(1)$.  The
baryon mass $M$ is $O(N_c)$, so the baryon acts as a heavy static source 
for the scattering of mesons at low energies.  
The absorption of the incoming
meson by the heavy baryon results in an intermediate baryon state which
is offshell by a four-momentum of order unity.
The momentum of an intermediate baryon
can be written as 
\begin{equation}
P^\mu = M v^\mu + k^\mu \equiv M v^{\prime \mu},
\end{equation} 
where $v^\mu$ is the four-velocity of the initial baryon and $k^\mu$ 
is a residual offshell momentum~\cite{bchpt}.   
The $M v^\mu$ piece of the momentum is $O(N_c)$ while the residual 
momentum is $O(1)$, so the intermediate baryon four-velocity
\begin{equation}
v^{\prime \mu} =  v^\mu + O\left( {1 \over N_c} \right)
\end{equation}
is equal to the initial baryon four-velocity in the large-$N_c$ limit,
and there is no recoil of the baryon.
In the large-$N_c$ limit, the incident and emitted mesons
have the same energy since no energy is transferred to the infinitely heavy
baryon, but the three-momentum of the meson changes in the scattering process.
The propagator of the intermediate baryon reduces to 
\begin{equation}
{{i \left( \slash P + M \right)} \over {P^2 - M^2}} \rightarrow
{i \over {k \cdot v} } \left( {{1 + \slash v} \over 2} \right),
\end{equation}
which does not involve the $O(N_c)$ baryon mass and is
manifestly $O(1)$.  In the rest frame of the baryon, the velocity
projection operator
projects out the two large components of the baryon Dirac spinor, and the
baryon propagator reduces to the nonrelativistic propagator $i/E$.
The baryon-meson scattering amplitude is $O(1)$ for any energy, but a
baryon-meson vertex grows as $O(\sqrt{N_c})$.  
Individual baryon-meson diagrams describing the absorption and then
emission of a meson by a baryon contain
two baryon-meson vertices, and grow
as $N_c$.  The total scattering amplitude grows as $N_c$ and violates
unitarity unless there are cancellations amongst the $O(N_c)$
baryon-meson 
scattering diagrams.  These cancellations are described by large-$N_c$ 
consistency conditions which constrain the baryon-meson couplings.
Large-$N_c$ consistency conditions can be derived for general meson 
couplings 
to baryons.  However, it is easiest to derive results for low-energy
pions because the leading order baryon-meson diagrams reduce to tree
diagrams~\cite{dm}.
A general analysis for an arbitrary meson involves a much larger class of
diagrams~\cite{mattis}. 

At low energies, a pion is derivatively coupled to baryons, and the $\pi + B
\rightarrow B^\prime$ vertex is described by 
\begin{equation}
{{{\partial_\mu} \pi^a} \over f_\pi } \left( A^{\mu a}\right)_{B^\prime B},
\end{equation}
where $f_\pi$ is the pion decay constant,
\begin{equation}
\left(A^{\mu a} \right)_{B^\prime B}
=\left< B^\prime |( \bar q \gamma^\mu \gamma_5 \tau^a q )_{\rm QCD} 
| B \right>
\end{equation}
is the isovector baryon axial vector current matrix element,
and $a=1,2,3$ is the isospin index of the pion.  The baryon-pion coupling
is $O(\sqrt{N_c})$ because the isovector baryon axial vector current
matrix element is $O(N_c)$ and 
$f_\pi \sim O(\sqrt{N_c})$.  In the large $N_c$ limit, the baryon is
infinitely heavy compared to the pion, so the baryon-pion coupling reduces
to the static baryon coupling
\begin{equation}
{{{\partial^i} \pi^a} \over f_\pi } \left( A^{i a}\right)_{B^\prime B}\ ,
\end{equation}
where $i=1,2,3$.  Notice that the
$p$-wave pion coupling has spin $1$ and isospin $1$.
It is useful to keep the $N_c$-dependence of the baryon axial vector current
manifest by defining 
\begin{equation}
A^{i a} \equiv g N_c X^{ia}\ , 
\end{equation}
where the baryon matrix elements
$X^{ia}$ are $O(1)$ and have a well-defined
large $N_c$ limit.  (It is to be understood in the following that $A^{ia}$ and
$X^{ia}$ refer to baryon matrix elements.)  
An $O(1)$ coupling constant $g$ is inserted in the definition so that a
convenient normalization for $X^{ia}$ can be chosen later on.

The amplitude for pion-baryon scattering at 
fixed pion energy $E$ is dominated in the large $N_c$ limit
by the diagrams displayed in Fig.~5.
The total amplitude of the two diagrams is proportional to
\begin{equation}\label{nxx}
N_c \left[ X^{ia}, X^{jb} \right],
\end{equation}
since Figs.~5(a) and~(b)
have the two
pion couplings in a different order and there is a relative
minus sign between the diagrams.  The propagator of the intermediate baryon
the intermediate
baryon in Fig.~5(a) is off-shell by an energy $E$ whereas the
intermediate baryon in Fig.~5(b) is off-shell by an energy $-E$.  
The product of the $X$'s in Eq. (\ref{nxx}) is a matrix product which 
sums over all possible baryon intermediate states. 
Only intermediate baryon states 
which are degenerate with the initial and final
baryons in the large $N_c$ limit contribute in the sum.  
Although each individual diagram in Fig.~5 contains
two baryon-pion vertices and is $O(N_c)$,
the total scattering amplitude
is at most $O(1)$, which leads to the constraint~\cite{dm}
\begin{equation}
N_c \left[ X^{ia}, X^{jb} \right] \le O(1)\ .
\end{equation}
The $1/N_c$ expansion for the operator $X^{ia}$ is given by
\begin{equation}\label{xexp}
X^{ia} = X_0^{ia} + {1 \over N_c} X_1^{ia} + {1 \over N_c^2} X_2^{ia} +
\ldots \ .
\end{equation}
The large-$N_c$ consistency condition implies that
\begin{equation}\label{xxcom}
\left[ X_0^{ia}, X_0^{jb} \right] = 0 ,
\end{equation}
which constrains baryon-pion couplings at leading order in the $1/N_c$ 
expansion.  Eq.~(\ref{xxcom}) was derived by Dashen and Manohar~\cite{dm}
and by Gervais and Sakita~\cite{gs}.

The solution of the large-$N_c$ consistency condition Eq.~(\ref{xxcom})
can be given in closed form~\cite{dm}.  
Assume that there is a baryon state with
$J=I=\frac 1 2$ which can be identified with the nucleon, and consider
the scattering process
$N + \pi \rightarrow N + \pi$.  If the only intermediate baryon
involved in the scattering is the nucleon, then the consistency condition
Eq.~(\ref{xxcom}) must be satisfied for the nucleon matrix elements of the
operator $X_0^{ia}$.
The nucleon matrix elements of $X_0^{ia}$ are
proportional to the nucleon matrix elements of $\sigma^i \tau^a$ 
by the Wigner-Eckart theorem.  Since the nucleon matrix elements of 
$\sigma^i \tau^a $
do not commute, Eq.~(\ref{xxcom}) is not satisfied and  
there must be additional baryon states
degenerate with the nucleon which participate as intermediate states in the
scattering.  The $p$-wave pion coupled to the nucleon can produce a baryon
state with $J=I= \frac 3 2$, which can be identified with the $\Delta$.  
Eq.~(\ref{xxcom}) for $N + \pi \rightarrow N + \pi$ scattering with
intermediate nucleon and $\Delta$ states yields an equation involving
$g_{\pi N \Delta}$ and $g_{\pi NN}$.  The large-$N_c$ consistency condition
determines $g_{\pi N \Delta}$ in terms of $g_{\pi NN}$.  
Next, consider $N + \pi \rightarrow \Delta + \pi$ scattering.  
The allowed intermediate states are $N$ and
$\Delta$, so Eq.~(\ref{xxcom}) yields an equation which determines 
$g_{\pi \Delta \Delta}$ 
in terms of $g_{\pi NN}$ and $g_{\pi N \Delta}$.  The consistency condition
for $\Delta + \pi \rightarrow \Delta + \pi$ scattering cannot be satisfied
unless there is an additional baryon state with $J=I= \frac 5 2$.  The
consistency condition for this scattering fixes the pion coupling of the
$\Delta$ state to the $J=I=\frac 5 2$ state in terms of the other couplings.
The recursion continues ad infinitum.  Thus, the solution of the large-$N_c$
consistency condition Eq.~(\ref{xxcom}) requires an infinite tower of
degenerate baryons with $I=J= \frac 1 2, \frac 3 2, \cdots$.  All of the 
pion couplings of these baryons are determined in terms of one overall
normalization $g$.  The matrix elements of $X_0^{ia}$ for the degenerate
baryon states $\left|J J_z, I I_z \right>$ are completely specified, and
are given by
\begin{equation}\label{x0}
\left< J^\prime J^\prime_z, I^\prime I^\prime_z \left| X_0^{ia} 
\right| J J_z, I I_z  \right>
= \sqrt{ {{2J+1} \over {2J^\prime+1}} }
\clebsch{I}{I_z}{1}{a}{I^\prime}{I^\prime_z}
\clebsch{J}{J_z}{1}{i}{J^\prime}{J^\prime_z}\ .
\end{equation}
The $p$-wave pions have diagonal couplings to each baryon and off-diagonal
couplings between baryons with spins and isospins differing by $J=I=1$. 
Thus, the degeneracy requirement for the baryon states is that
the mass splittings of baryons between adjacent baryons in the spin tower
are at most order $1/N_c$.
The solution Eq.~(\ref{x0}) of the large-$N_c$ consistency condition 
implies that the ratios of all pion-baryon
couplings are determined in the large-$N_c$ limit.  
For example, one finds
that
\begin{equation}
{ g_{\pi \Delta N} \over g_{\pi NN}} = \frac 3 2, 
\end{equation} 
which is the model-independent result derived previously in the
context of the Skyrme model~\cite{anw}.  The ratio $g_{\pi \Delta \Delta}/g_{\pi
NN}$ also is determined.  
The overall normalization $g$ of the couplings, however,
is not determined by the large-$N_c$ constraint.

\subsection{\it Contracted Spin-Flavor Algebra}

The contracted spin-flavor algebra for large-$N_c$
baryons is given by the large-$N_c$
consistency condition
\begin{equation}\label{xx}
\left[ X_0^{ia}, X_0^{jb} \right] = 0 ,
\end{equation}
the spin and flavor algebras
\begin{equation}\label{sfalg}
\left[J^i, J^j \right] = i \epsilon^{ijk} J^k, 
\quad\left[I^a, I^b \right] = i \epsilon^{abc} I^c,
\quad\left[J^i, I^a \right] = 0, 
\end{equation}
and the commutation relations 
\begin{equation}\label{jxix}
\left[J^i, X_0^{ja} \right] = i \epsilon^{ijk} X_0^{ka}, \qquad
\left[I^a, X_0^{ib} \right] = i \epsilon^{abc} X_0^{ic} \ . 
\end{equation}
The commutators Eq.~(\ref{jxix}) follow 
because $X_0^{ia}$ transforms under
spin and isospin as an irreducible tensor operator with spin $1$ and 
isospin $1$.

It is useful to compare the contracted $SU(4)$ spin-flavor algebra
Eqs.~(\ref{xx})-(\ref{jxix}) to the $SU(4)$ spin-flavor algebra.
The $SU(4)$ generators $J^i$, $I^a$ and $G^{ia}$ satisfy the spin-flavor
algebra Eq.~(\ref{sfalg}) and
\begin{eqnarray}
&&\left[J^i, G^{ja} \right] = i \epsilon^{ijk} G^{ka}, \qquad
\left[I^a, G^{ib} \right] = i \epsilon^{abc} G^{ic}, \nonumber\\
&&\left[G^{ia}, G^{jb} \right] = 
{i \over 4} \delta^{ab}\epsilon^{ijk} J^k +
{i \over 4} \delta^{ij}\epsilon^{abc} I^c\ . 
\end{eqnarray}
The spin-flavor generator $G^{ia}$ with $J=I=1$
has matrix elements which are $O(N_c)$ for
large-$N_c$ baryons, so $G^{ia}/N_c$ has a nonvanishing large-$N_c$ limit.
The large-$N_c$ contracted $SU(4)$ spin-flavor algebra for baryons is obtained
from the $SU(4)$ algebra by rescaling the spin-flavor generator $G^{ia}$ by a
factor of $1/N_c$ and taking the large $N_c$ limit,
\begin{equation}
\lim_{N_c \rightarrow\infty}{ G^{ia} \over N_c} \rightarrow X_0^{ia} \ .
\end{equation}
The spin $\otimes$ flavor algebra is unaffected by this contraction, and 
the commutators which are homogeneous in $G^{ia}$ yield the commutators
Eq.~(\ref{jxix}).  The commutation relation for $[ G^{ia}, G^{jb} ]$ is
divided by two powers of $N_c$, and reproduces the large-$N_c$ consistency
condition Eq.~(\ref{xx}) in the limit $N_c \rightarrow \infty$, since baryon
matrix elements of $J$ and $I$ are at most $O(N_c)$. 

The contracted spin-flavor algebra for large-$N_c$ baryons arises because the
axial vector current for large-$N_c$ baryons grows with $N_c$.  The rescaled
axial vector current $X_0^{ia}$ has a well-defined large-$N_c$ limit, and
can be treated like a classical commuting object.  The usual spin and flavor
algebra of baryons is extended to a spin-flavor algebra in the large-$N_c$
limit when $X_0^{ia}$ is included in the algebra.  
Note that a spin-flavor symmetry which
mixes internal and Lorentz symmetries can emerge for large-$N_c$ baryons
because the baryon field is static in the large $N_c$ limit.

\subsection{\it Large-$N_c$ Baryon Representations}

All possible irreducible representations of the contracted spin-flavor algebra
have been classified
using the theory of induced representations in Ref.~\cite{djm2}.  
The eigenvalues of the commuting generator $X_0^{ia}$ 
can be used to label states.
A standard
reference state with coordinate $X_0^{ia}$ diagonalized can be chosen for each
irreducible representation.  The reference state can be
regarded as the analogue of the highest weight state of an irreducible
representation of a Lie algebra.  All states in an irreducible representation
are obtained from the reference state by applying group
transformations, just as all states in an irreducible representation can be
obtained from the highest weight state by making group transformations.  The
abelian coordinate $X_0^{ia}$ is not affected by the action of the generator
$X_0$, so all
states are obtained by spin and isospin transformations of the
reference state.  The standard reference state for irreducible
representations corresponding to large $N_c$ baryons is
\begin{equation}
X_0^{ia} = \delta^{ia}={\rm diag} (1,1,1),
\end{equation}
where $i=1,2,3$ and $a=1,2,3$ label the rows and columns of
the $3 \times 3$ matrix.  This reference state is invariant
under transformations of a $SU(2)$ little group
generated by
\begin{equation}
{\bf K} = {\bf I} + {\bf J} ,
\end{equation}
so a representation of the little group must be specified to completely 
label the states.  The states $\left| X_0^{ia}, K, k \right>$ of an
irreducible representation are labeled by the abelian coordinate 
$X_0^{ia}$ and representations $\left| K k \right>$ of 
the $SU(2)$ little group.

The basis states $\left| X_0^{ia}, K, k \right>$ of an irreducible 
representation diagonalize the baryon axial vector currents, 
but are not states of definite spin and isospin.
Baryons are states of definite spin and
isospin, so it is necessary to project the basis states 
$\left| X_0^{ia}, K, k \right>$ onto states
which diagonalize spin and isospin.  These new basis states 
$\left| I I_z, J J_z; K \right>$ can be identified with baryons, but now
the states
do not diagonalize the baryon axial vector current.  Thus,
the baryon axial vector current has off-diagonal matrix elements
connecting baryon states with different spin and isospin.

Large-$N_c$ baryon representations contain
all spin and isospin representations 
$(J,I)$ which are consistent with a given value of $K$.
For the physical case of $N_c$ odd,
the lowest allowed baryon spin is $J = \frac 1 2$.
The irreducible representation with $K=0$ is given by the infinite tower of
$(J,I)$ states 
\begin{equation}\label{jeqi}
\left(\frac 1 2, \frac 1 2 \right), 
\left(\frac 3 2, \frac 3 2\right), \left(\frac 5 2, \frac 5 2 \right),
\ldots\ .
\end{equation}
The irreducible representation with $K= \frac 1 2$ corresponds to the infinite
tower of states
\begin{equation} 
\left(\frac 1 2, 0 \right), \left(\frac 1 2, 1 \right),
\left(\frac 3 2, 1 \right), \left(\frac 3 2, 2 \right),
\left(\frac 5 2, 2 \right), \left(\frac 5 2, 3 \right), \ldots \ .
\end{equation}
The irreducible representation with $K=1$ is the infinite tower
\begin{equation} 
\left(\frac 1 2, \frac 1 2 \right), 
\left(\frac 1 2, \frac 3 2 \right), 
\left(\frac 3 2, \frac 1 2\right), \left(\frac 3 2, \frac 3 2 \right),
\left(\frac 3 2, \frac 5 2 \right),
\left(\frac 5 2, \frac 3 2 \right), \left(\frac 5 2, \frac 5 2 \right), 
\left(\frac 5 2, \frac 7 2 \right), 
\ldots \ ,
\end{equation}
while the irreducible representation for $K= \frac 3 2$ is the infinite tower
\begin{equation} 
\left(\frac 1 2, 1 \right), \left(\frac 1 2, 2 \right),
\left(\frac 3 2, 0 \right), \left(\frac 3 2, 1 \right),
\left(\frac 3 2, 2 \right), \left(\frac 3 2, 3 \right), \ldots \ .
\end{equation}
The low-spin states of the different $K$ towers have the correct spin and
isospin quantum numbers to be identified with the spin-$\frac 1 2$ octet
and spin-$\frac 3 2$ decuplet baryon states of QCD if the quantum number $K$ is
related to the number of strange quarks in a baryon, $K = N_s/2$.  The $K=0$
tower contains strangeness zero baryons such as the nucleon state $(\frac 1 2,
\frac 1 2)$ and the $\Delta$ state $(\frac 3 2, \frac 3 2)$.  The $K = \frac 1
2$ tower contains strangeness $-1$ baryons 
$\Lambda(\frac 1 2,0)$, $\Sigma(\frac 1 2, 1)$ and $\Sigma^*(\frac 3 2, 1)$.
The $K=1$ tower contains the strangeness $-2$ baryons 
$\Xi(\frac 1 2,\frac 1 2)$ and $\Xi^*(\frac 3 2,\frac 1 2)$, 
while the $K= \frac 3 2$ tower contains
the strangeness $-3$ baryon $\Omega(\frac 3 2,0)$.
The other states in the large-$N_c$ towers correspond to baryons which exist
for $N_c \rightarrow \infty$, but not for $N_c=3$.  Notice that the 
towers with $K >0$
contain extra baryon states even for $J= \frac 1 2$ and $J=\frac 3 2$.

For large and finite $N_c$, one can work with the  
$SU(2N_F)$ spin-flavor algebra with $N_F$ light quark flavors rather than
the contracted spin-flavor algebra.
The lowest-lying
baryon spin-flavor representation is given by the completely symmetric 
$SU(2N_F)$ representation
with $N_c$ boxes shown in Fig.~6.
Under the decomposition
$SU(2N_F) \rightarrow SU(2) \otimes SU(N_F)$, the spin-flavor representation
yields the tower of spin and flavor representations shown in
Fig.~7.
For $N_F=2$, the ground state baryon representation
produces the tower of baryon states 
\begin{equation}\label{finitetower}
J=I=\frac 1 2, \frac 3 2, \ldots , \frac {N_c} 2 \ .
\end{equation}
The baryon tower is now finite dimensional,
unlike the baryon tower Eq.~(\ref{jeqi})
of the contracted spin-flavor algebra.
The baryon tower Eq.~(\ref{finitetower})
reduces to the states $N$ and $\Delta$ for $N_c =3$. 

It is instructive to consider how the quark spin and flavor is arranged in the
large-$N_c$ baryon.  The large-$N_c$ nucleon with $J=I=\frac 1 2$
is the tensor product of $\nu= (N_c-1)/2$ up and down
quarks arranged in a spin and flavor singlet
\begin{equation}
\frac 1 2 \left|\left(u d - d u \right) \times 
\left( \uparrow \downarrow - \downarrow \uparrow \right)\right>
\end{equation}
and one additional up or down quark with spin up or down.  The spin and
isospin quantum numbers of the nucleon are determined by this one additional
quark.  For example, 
the large-$N_c$ proton state is the tensor product of 
$\nu$ down quarks and $(\nu + 1)$ up quarks.
It is clear that the low-spin baryon wavefunctions are highly ordered in the
spin and flavor wavefunctions of the $N_c$ quarks.
This ordering can be described by a collective coordinate.
In a quark model, the baryon annihilation operator can be written
as
\begin{equation}
q_{\imath_1 \alpha_1}q_{\imath_2 \alpha_2}\cdots q_{\imath_N \alpha_N}
A_{\imath_1 \alpha_1}A_{\imath_2 \alpha_2}\cdots A_{\imath_N \alpha_N}, 
\end{equation}
where the collective coordinate
$A$ is a $2 \times 2$ matrix with rows and columns labeled by spin and
isospin.
The states of the Skyrme model also are labeled by a collective coordinate.
The large-$N_c$ basis states $\left| X_0^{ia}, K, k \right>$ are identical to the
Skyrme model states with the coordinate $X_0^{ia}$ related 
to the collective coordinate $A$ by
\begin{equation}
X_0^{ia} = \frac 1 2 {\rm Tr} \left( A \sigma^i A^{-1} \tau^a \right).
\end{equation}
Notice that the operator $X_0^{ia}$ is completely symmetric in 
its vector and isovector
indices $i$ and $a$, and that the soliton occurs in
a $2 \times 2$ subspace.

\subsection{\it $1/N_c$ Corrections}

Large-$N_c$ consistency conditions also
can be derived for the $1/N_c$ corrections
of static baryon matrix elements~\cite{dm,j,djm1}.  
The $1/N_c$ corrections to the baryon mass
and axial couplings are described here.  The general form of the baryon
$1/N_c$ expansion is discussed as well. 

\subsubsection{Masses: No $1/N_c$ Correction in Flavor Symmetry Limit}

Consistency conditions are derived for the baryon mass by
considering the effect of a baryon mass splitting 
on the pion-baryon scattering amplitude.  The intermediate baryon propagator
for nondegenerate baryons is given by $i/(k\cdot v - \Delta M)$, where
$\Delta M$ is a baryon mass splitting.  Expanding this propagator to first
order in a power series in $\left(\Delta M/ (k \cdot v)\right)$ 
yields the constraint
\begin{equation}\label{mcc}
\left[ X^{ia}, \left[ X^{jb}, M \right] \right] \le 
O\left( { 1 \over N_c} \right)\ .
\end{equation}
The baryon mass $M$ has an expansion in $1/N_c$ of the form
\begin{equation}
M = N_c M_0 + M_1 + {1 \over N_c} M_2 + \ldots \ .
\end{equation}
Eq.~(\ref{mcc}) implies that $M_0$ and $M_1$ satisfy
\begin{equation}
\left[ X_0^{ia}, \left[ X_0^{jb}, M_0 \right] \right]=0, \qquad
\left[ X_0^{ia}, \left[ X_0^{jb}, M_1 \right] \right]=0 \ .
\end{equation}
The only solution of these large-$N_c$ consistency conditions 
is $M_0 = M_1 = \openone$, so the baryon tower is
degenerate upto $1/N_c$ corrections as expected,
\begin{equation}
M = m_0 N_c \openone + O\left( {1 \over N_c} \right)\ .
\end{equation}
The $1/N_c$ correction to the baryon mass is constrained by considering
the scattering pion + baryon $\rightarrow$ baryon + 2 pions, which
yields the constraint
\begin{equation}
\left[ X^{ia}, \left[ X^{jb}, \left[ X^{kc}, M \right] \right] \right] 
\le O\left( { 1 \over {N_c^2} } \right)\ .
\end{equation}
This constraint yields a large-$N_c$ consistency condition for $M_2$,
\begin{equation}
\left[ X_0^{ia}, \left[ X_0^{jb}, \left[ X_0^{kc}, M_2 \right] \right] \right]
=0, 
\end{equation}
which
has solutions $M_2=J^2$, $I^2$ and $X_0^{ia} X_0^{ia}$.  The operator $I^2 = J^2$
and the operator $X_0^{ia} X_0^{ia}=3$, so the only independent solution is
$M_2 = J^2$.
Thus, the baryon mass is
given by
\begin{equation}
M = m_0 N_c \openone + m_2 {1 \over N_c } J^2 + \ldots \ .
\end{equation}

The hyperfine mass splittings 
of the $J=I= \frac 1 2, \frac 3 2, \cdots$ baryon tower are depicted in
Fig.~8.
For low-spin baryons with $J \sim O(1)$, the
hyperfine splitting $J^2/N_c$ is order $1/N_c$, whereas for baryons with
$J \sim O(N_c)$, the hyperfine splitting is order $N_c$ with respect to the
low-spin baryons.  The $1/N_c$ expansion is under control for baryons with
fixed and finite spin and isospin when the large-$N_c$ limit is taken, but
the expansion breaks down for baryons with spin and isospin $O(N_c)$.  

The mass spectrum of the baryon tower with $K=0$ can be related to the mass
spectra of baryon towers with different $K$ by considering kaon-baryon
scattering~\cite{djm1}.  The masses of the different $K$ towers are related by
\begin{equation}
M = m_0 N_c \openone + m_1 K + \ldots \ , 
\end{equation}
so the baryon towers with differing strangeness
are not degenerate, and the leading mass splitting between different baryon 
towers is linear in strangeness.  Higher order corrections also can be 
classified.

\subsubsection{Axial Couplings: No $1/N_c$ Correction for $N_F =2$}

A consistency condition for the $1/N_c$ correction to the baryon axial vector
current is obtained by considering the scattering process baryon + pion
$\rightarrow$ baryon + 2 pions, shown in Fig.~9.
Each baryon-pion
vertex is $O(\sqrt{N_c})$, so individual diagrams are $O(N_c^{3/2})$ but the
total amplitude is at most $O(1/\sqrt{N_c})$.  Summing over all the tree
diagrams yields the constraint~\cite{dm}
\begin{equation}
N_c^{3/2} \left[ X^{ia}, \left[ X^{jb} , X^{kc} \right] \right]  \le
O\left({1 \over {\sqrt{N_c}}}\right) \ ,
\end{equation}
which implies that the double commutator of three $X$'s vanishes at least as
fast as $O(1/N_c^2)$,
\begin{equation}
\left[ X^{ia}, \left[ X^{jb} , X^{kc} \right] \right] \le 
O\left( {1 \over N_c^2 } \right) \ .
\end{equation}
Expanding $X^{ia}$ in $1/N_c$ gives
the large-$N_c$ consistency condition for $X_1^{ia}$
\begin{equation}
\left[X_0^{ia}, \left[ X_0^{jb} , X_1^{kc} \right] \right] +
\left[X_0^{ia}, \left[ X_1^{jb} , X_0^{kc} \right] \right] =0\ .
\end{equation}
The unique solution to this consistency condition is 
$X_1^{ia} \propto X_0^{ia}$.  There is no other candidate operator since
the operators $\epsilon^{abc}I^b X_0^{ic}$
and $\epsilon^{ijk} J^j X_0^{ka}$ do not transform in 
the same manner as the axial current under time reversal, and the operator
$\epsilon^{ijk} \epsilon^{abc} X_0^{jb} X_0^{kc}$ is equal to $2 X_0^{ia}$.  
Eq.~(\ref{xexp}) implies that $X^{ia}$ is equal to $X_0^{ia}$ times $(1 + c/N_c)$
where $c$ is an arbitrary coefficient, upto corrections of order $1/N_c^2$.
The $1/N_c$ proportionality factor can be
absorbed into the unknown coupling $g$, so
\begin{equation}
X^{ia} = X_0^{ia} + O\left( {1 \over N_c^2} \right)\ , 
\end{equation}
and all ratios of pion-baryon couplings are determined as before
upto corrections
of $O(1/N_c^2)$.

The above derivation is valid for any baryon spin tower
with fixed $K$. 
The normalization
of the baryon axial vector currents for different spin towers with differing
strangeness can be related by considering kaon-baryon scattering~\cite{djm1}. 
The result
is 
\begin{eqnarray}
&&\left< J^\prime J^\prime_z, I^\prime I^\prime_z; K \left| A^{ia} 
\right| J J_z, I I_z ; K \right>
= N_c g(K) (-1)^{2J^\prime+J-I^\prime-K}\times \nonumber\\
&&\qquad \sqrt{ (2I+1) (2J+1) }
\sixj{1}{I}{I^\prime}{K}{J^\prime}{J}
\clebsch{I}{I_z}{1}{a}{I^\prime}{I^\prime_z}
\clebsch{J}{J_z}{1}{i}{J^\prime}{J^\prime_z} \nonumber
\end{eqnarray}
with 
the couplings of different $K$ towers 
related by 
$g(K) = g(0) + g_1 K/N_c + \ldots$,
where $g_1$ is an arbitrary coefficient.  The value of $g_1$ can be determined
in the $SU(3)$ flavor symmetry limit~\cite{djm1}.  The leading correction to the
coupling constants $g(K)$ of different $K$ towers is linear in strangeness, so
there is an equal spacing rule for the pion-baryon couplings of baryons with
different strangeness~\cite{djm1}.

Higher order $1/N_c$ corrections to the baryon axial vector current can be
studied by considering baryon-pion scatterings to a baryon + $n$ pions, where
$n>2$.
The first nontrivial
correction $X_2^{ia} = J^i I^a$ occurs at $O(1/N_c^2)$, 
so the large-$N_c$ values of baryon-pion coupling
ratios should be violated at this level.  The measured couplings are observed 
to satisfy the relations at the $10\%$ level, which is what one would
naively expect for $1/N_c = 1/3$. 

\subsubsection{General Form of $1/N_c$ Expansion}

It is possible to write down the $1/N_c$ expansion for a baryon operator
directly without deriving the large-$N_c$ consistency conditions.
The $1/N_c$ expansion for any static baryon operator for $N_F=2$ light quark
flavors is of the form 
\begin{equation}
N_c {\cal P}\left( X_0, {J \over N_c}, {I \over N_c} \right),
\end{equation}
where ${\cal P}$ is a polynomial in its arguments.
It is to be understood that all possible independent operator products 
constructed out of the generators
of the contracted spin-flavor algebra appear in the $1/N_c$ expansion with
arbitrary coefficients.  
The generator $X_0^{ia}$ has matrix elements of order
unity for all baryons in a given irreducible representation of the spin-flavor
algebra.  The parameters $J/N_c$ and $I/N_c$ are $O(1/N_c)$ for baryons with
spin and isospin of order unity at the bottom of the baryon towers.  For these
baryons, operators with more powers of $J/N_c$ and $I/N_c$ are suppressed
in $1/N_c$ compared to operators with fewer powers.  Baryons will $J \sim
O(N_c)$ at the top of the baryon tower, have $J/N_c$ and $I/N_c$ of order
$1$, so all terms in the $1/N_c$ expansion are equally important, and the
$1/N_c$ expansion is not predictive.  
The $1/N_c$
expansion can be extended to baryon towers with different $K$ by including
additional operators such as ${\bf K} / N_c$ in the polynomial 
${\cal P}$.  This extension gives the form of the $1/N_c$ expansion for
$N_F = 3$ light quark flavors with arbitrary $SU(3)$ breaking because only
isospin flavor symmetry is imposed.

\section{$1/N_c$ EXPANSION}

The spin-flavor symmetry of large-$N_c$ baryons provides the 
organizational framework for the baryon $1/N_c$ expansion.     
%
%
%
As we have seen,
the $1/N_c$ expansion of any static baryon operator is given in terms of 
operator products of the baryon spin-flavor generators.
Not all operator products are linearly
independent, so redundant operators must be
eliminated from the expansion using operator identities.  
Each independent operator product appears at a 
definite order in $1/N_c$,
accompanied by an uncalculable coefficient.  
The $1/N_c$ expansion is predictive for baryons with finite spin and
flavor quantum numbers because only a finite number
of operator products occur at any given order in the $1/N_c$ expansion.

There are two natural operator realizations of the baryon spin-flavor algebra 
which can be used to construct the operator basis of the $1/N_c$ expansion.
The first uses the spin-flavor generator $X_0^{ia}$ in operator
products, whereas the second realization uses the $SU(2N_F)$ 
commutation relations with spin-flavor generator $G^{ia}$.
The operators $X_0^{ia}$ and $G^{ia}/N_c$ differ at subleading order $1/N_c$.
The ambiguity in the choice of spin-flavor generator arises because
the contracted spin-flavor algebra for baryons is exact only in the 
large $N_c$ limit, so the matrix elements of the spin-flavor generators which
complete the spin-flavor algebra 
are known only to leading order in $1/N_c$ upto a normalization factor.  
When $X_0^{ia}$ is used, the operator basis of the $1/N_c$
expansion is the same as the operator basis of the large-$N_c$ Skyrme model,
whereas the operator basis constructed using 
$G^{ia}$ is the same as the operator basis of the large-$N_c$ nonrelativistic 
quark model.  
Both operator bases parametrize the same large-$N_c$ physics
since all operator products appear in the $1/N_c$ expansion with uncalculated
coefficients.  In principle, these coefficients are calculable in terms of the
complicated QCD dynamics, but in practice the coefficients are treated as
unknowns since any such calculation would be tantamount to solving QCD exactly
in the baryon sector.  The coefficients of the $1/N_c$ expansion
for large-$N_c$ QCD baryons are different from the coefficients of the 
large-$N_c$ Skyrme model and the large-$N_c$ quark model, but group theoretic
predictions which do not depend on the coefficients will be the same for
large-$N_c$ QCD and these models.  

The general form of the $1/N_c$ expansion using $X_0^{ia}$ was explained 
briefly in the previous section.  The solution was given for three light quark
flavors with exact isospin symmetry and arbitrary $SU(3)$ flavor breaking.
The analysis using $G^{ia}$ will be performed in this section.  The discussion
is first restricted to the case of two light quark flavors with exact isospin
flavor symmetry.  The generalization to $N_F=3$ light quark flavors with
$SU(3)$ flavor symmetry breaking is presented later in the section.

\subsection{\it Baryon Operator Expansion}

The static baryon matrix elements of a QCD $1$-body operator,
such as $(\bar q \gamma^\mu \gamma_5 \tau^a q)_{\rm QCD}$, have a $1/N_c$
expansion of the form
\begin{equation}\label{opexp}
{\cal O}^{\rm 1-body}_{\rm QCD} 
= N_c \sum_{n} c_n { 1 \over {N_c^{n} }} {\cal O}_n \ ,
\end{equation}
where the sum is over all independent operator polynomials ${\cal O}_n$ of
degree $n$ in the baryon
spin-flavor generators which transform 
according to the same spin $\times$ flavor representation
as ${\cal O}_{\rm QCD}$.  Here spin refers to the $SU(2)$ algebra 
generated by the baryon spin $J^i$ in the rest frame of the baryon.
For finite $N_c$, the sum on $n$ is over $0 \le n \le N_c$.
Every operator coefficient $c_n(1/N_c)$ has an expansion in $1/N_c$ 
beginning at order unity.  The factor $1/N_c^n$ is present since each
spin-flavor generator in an $n^{\rm th}$ order 
operator product ${\cal O}_n$ is accompanied by a
factor of $1/N_c$.  The overall factor of
$N_c$ arises because a QCD $1$-body operator 
has matrix elements which
are at most $O(N_c)$ when inserted on
all quark lines of the baryon.  
The generalization of the $1/N_c$ expansion to a QCD 
$m$-body operator acting on $m$ quarks 
is obtained by replacing $N_c= N_c^1$ by $N_c^{m}$ in Eq.~(\ref{opexp}).  

The baryon $1/N_c$ expansion depends on $N_c$ through the explicit
$1/N_c$ prefactors and through the implicit $N_c$-dependence of the
${\cal O}_n$ matrix elements.  The matrix elements of the operators
${\cal O}_n$ have a nontrivial dependence on $N_c$ which varies
for different states of the baryon spin-flavor multiplet.   
This dependence on $N_c$ can be obtained by analyzing
the $N_c$-dependence of the baryon spin-flavor generators.  
The baryon spin tower $J=I=\frac 1 2, \frac 3 2, \ldots, \frac {N_c} 2$
contains baryons at the bottom of the spin tower
with spin $J$ and isospin $I$ of order unity, as well as baryons at the top
of the spin tower with spin and isospin of $O(N_c)$.  All of the baryon
states in the
multiplet have matrix elements of $G^{ia}$ that are $O(N_c)$.
Since the matrix elements of $J$, $I$ and $G$ are $\le O(N_c)$ on the baryon
states,
$n^{\rm th}$ order operator polynomials ${\cal O}_n$ constructed from the
generators have matrix elements which are $\le O(N_c^n)$. 
Baryon states with $J \sim O(N_c)$ have ${\cal O}_n$ matrix elements which are
$O(N_c^n)$, so the
explicit factor of $1/N_c^n$ is completely cancelled by the implicit 
$N_c^n$ dependence of the operator matrix elements.  In this case, all
terms in the $1/N_c$ expansion are equally important and must be retained at
leading order, so the $1/N_c$ expansion is not
predictive. 
For baryon states with $J \sim O(1)$,
operator polynomials containing more factors of $J$ are
systematically suppressed in $1/N_c$, and the operator expansion
can be truncated at various orders in the $1/N_c$ expansion.  The $1/N_c$
expansion for the low-spin baryons is predictive because not all spin-flavor
structures appear at a given order in $1/N_c$.

The group theoretic structure of the $1/N_c$
expansion is determined by the operators ${\cal O}_n$ and the explicit
$1/N_c$ prefactors.        
The complicated large-$N_c$ QCD dynamics 
is parametrized by the unknown coefficients $c_n(1/N_c)$.  The quark-gluon
diagrams of large-$N_c$ QCD contribute to the $c_n(1/N_c)$ at different orders
in $1/N_c$ depending on their topological structure.   
Diagrams of $N_c$ valence quark
lines with arbitrary planar gluon exchange contribute to the coefficients
at leading order in $1/N_c$.  Diagrams containing quark loops and nonplanar
gluons contribute to the coefficients at subleading orders.  Each quark loop in
a baryon diagram is suppressed by one factor of $1/N_c$, and each nonplanar
gluon exchange is suppressed by $1/N_c^2$.  

The operators ${\cal O}_n$ which form the operator basis of the 
$1/N_c$ expansion have a natural interpretation as $n$-body quark operators
acting on the spin and flavor indices of $n$ static quarks.  This identification
is possible because the spatial dependence of the quark wavefunctions
is irrelevant for the computation of static baryon
matrix elements of the ground state baryons, so it is only necessary to
keep track of the spin, flavor and color quantum numbers of the quarks.  
(This point will be discussed in greater detail later in the context of 
the Hartree picture of large-$N_c$ baryons.)
A large-$N_c$ baryon state is totally antisymmetric in the
color indices of its $N_c$ constituent valence quarks, so the color structure of
the large-$N_c$ baryon is trivial, and color indices of the quarks 
also can be omitted.  
When the color indices of the quarks are omitted, the quarks
are treated as bosonic objects, since it is the
the antisymmetry of the $SU(N_c)$ color $\epsilon$-symbol combined with the Fermi
statistics of the quarks which implies that the ground state baryons contain 
$N_c$
quarks in the completely symmetric representation of spin $\otimes$ flavor.
Define a set of quark creation and 
annihilation operators, $q^\dagger_{\imath \alpha}$ and $q^{\imath \alpha}$, 
where $\imath=1,2$ and $\alpha=1, \ldots, N_F$ represent the spin and flavor 
quantum numbers of the quarks.  The creation and annihilation operators 
satisfy the bosonic commutation relation   
\begin{equation} 
\left[q^{\imath\alpha},q^\dagger_{\jmath\beta}\right] 
= \delta^\imath_\jmath\delta^\alpha_\beta .  
\end{equation}
An $n$-body quark operator is constructed in terms of 
these static quark creation and annihilation operators.

A systematic classification of $n$-body quark operators is possible.
There is a unique $0$-body operator $\openone$, which is the identity
operator acting on baryons.  This operator does not act on the quarks 
in the baryon.   
The $1$-body operators that act on a single quark consist of the
quark number operator $q^\dagger q$ and
\begin{eqnarray}
&&J^i = q^\dagger \left({ \sigma^i \over 2} \otimes \openone \right) q,
\nonumber \\
&&I^a = q^\dagger \left(\openone \otimes {\tau^a \over 2} \right) q,\\
&&G^{ia} = q^\dagger \left({\sigma^i \over 2} 
\otimes {\tau^a \over 2}\right) q, \nonumber
\end{eqnarray}
where $\sigma^i$ and $\tau^a$
are the $SU(2)$ spin and isospin generators acting on the spin-$\frac 1 2$
and isospin-$\frac 1 2$ indices of the quark creation and annihilation
operators.  The last three $1$-body quark operators can be identified with the
baryon spin-flavor generators $J^i$, $I^a$ and $G^{ia}$ when acting on 
baryon states, since the baryon matrix elements of
a $1$-body quark operator are obtained by summing over all possible   
insertions of the operator on the $N_c$ quark lines
of the baryon.  Thus, $J^i$, the $SU(2)$ baryon spin generator, is
equal to the sum of the spins of the $N_c$ quarks in the baryon,  
\begin{equation}
J^i = \sum_{\ell=1}^{N_c} q_\ell^\dagger \left({\sigma^i \over 2} 
\otimes {\openone}\right) q_\ell \ ,
\end{equation}
where $\ell$ denotes the quark line number.  Similarly, the baryon isospin
generator $I^a$ is given by the sum of the isospins of the $N_c$ quarks,
while the spin-flavor generator $G^{ia}$ is obtained by inserting both
$\sigma^i/2$ and $\tau^a/2$ on each quark line,
\begin{equation}
G^{ia} = \sum_{\ell=1}^{N_c} q_\ell^\dagger \left({\sigma^i \over 2} 
\otimes {\tau^a \over 2}\right) q_\ell \ .
\end{equation}
Inserting the quark number operator on each quark line of the baryon counts the
number of quarks in the baryon,
\begin{equation}
q^\dagger q= \sum_{\ell =1}^{N_c} q_\ell^\dagger q_\ell = N_c \openone\ .
\end{equation}
Thus, baryon matrix elements of the quark number operator 
can be rewritten in terms of the $0$-body baryon identity operator $\openone$,
and the only independent $1$-body operators are $J^i$, $I^a$ and $G^{ia}$.
The $n$-body operators for $n \ge 2$ 
can be written as completely
symmetric operator products of degree $n$ in the baryon spin-flavor generators.
An example of a $2$-body operator is
\begin{equation}
J^i T^a= \sum_{\ell, \ell^\prime} 
\left( q^\dagger_{\ell} {\sigma^i \over 2} q_{\ell} \right)
\ \left( q^\dagger_{\ell^\prime} {\tau^a \over 2} q_{\ell^\prime} \right) ,
\end{equation} 
which differs from the $1$-body operator 
$G^{ia}$ because $\sigma^i/2$ and $\tau^a/2$ are inserted independently 
on the quark lines.
Only symmetric products (anticommutators) of baryon spin-flavor generators
occur because antisymmetric products (commutators) reduce to linear
combinations of lower-body operators by the $SU(2N_F)$ Lie algebra commutation 
relations.
Not all 
of the $n^{\rm th}$ order operator products correspond
to independent operators.
The classification of the independent $n$-body operator polynomials 
for $n \ge 2$ is
nontrivial, and requires additional discussion.

It is convenient to consider normal ordered $n$-body operators
\begin{equation}\label{normalordered}
 q^\dagger_{\jmath_1 \beta_1} \ldots q^\dagger_{\jmath_n \beta_n}\  
{\cal T}_{(\imath_1 \alpha_1) \cdots (\imath_n \alpha_n)}^{(\jmath_1 \beta_1) 
\cdots (\jmath_n \beta_n)}\  
q^{\imath_1 \alpha_1} \ldots q^{\imath_n \alpha_n}\ ,
\end{equation}
where ${\cal T}$ denotes a spin and flavor tensor which saturates the
spin and flavor indices of the $n$ quark creation and annihilation operators.
The 
order of the $n$ creation operators or the $n$ annihilation operators 
in the normal ordered operator is unimportant, since the quark creation 
and annihilation operators are bosonic operators.
Normal ordered operators always act on $n$ distinct quarks because
$n$ quarks are first annihilated and then created.
The independent $n$-body operators are all normal-ordered $n$-body operators
with traceless and completely symmetrized tensors ${\cal T}$.  
The tensor ${\cal T}$ must be completely 
symmetrized in its upper and lower pairs of 
spin and flavor indices because
the quarks in the ground state baryons are in the completely symmetric
representation of the spin-flavor group, so tensor operators with mixed symmetry
vanish identically when acting on the ground state baryons.  The tensor   
${\cal T}$ is also traceless, since all nonvanishing 
traces of the tensor ${\cal T}$ reduce to lower-body operators.  

It is easy to identify the independent $n$-body operators when the operators
are written in normal ordered form, but normal
ordered operators do not have simple baryon matrix elements. 
For example, the normal ordered $2$-body operator
\begin{equation}
q^\dagger_{\jmath_1 \alpha_1} q^\dagger_{\imath_2 \beta_2}
\left({{\sigma^i} \over 2}\right)^{\jmath_1}_{\imath_1}
\left({{\tau^a} \over 2}\right)^{\beta_2}_{\alpha_2}
q^{\imath_1 \alpha_1} q^{\imath_2 \alpha_2} =
\sum_{\ell \neq \ell^\prime} 
\left( q^\dagger_{\ell} {\sigma^i \over 2} q_{\ell} \right)
\ \left( q^\dagger_{\ell^\prime} {\tau^a \over 2} q_{\ell^\prime}\right),
\end{equation}
does not equal 
$J^i T^a$ since the double sum over quark lines is restricted to pairs of 
distinct quark lines $\ell \neq \ell^\prime$.

In contrast to the situation for normal ordered operators,
the baryon matrix elements of operator products
of the baryon spin-flavor generators
are simple to evaluate, but the classification
of independent operator products is nontrivial.  When operator products
are rewritten in normal ordered form, 
there are tensors ${\cal T}$ which are not
traceless or completely symmetrized, so some
linear combinations of the operator products are equivalent to lower body 
operators or vanish identically on the completely symmetric ground state 
baryon representation.  Operator identities describe   
the operator product combinations which are redundant operators
and must be eliminated from the operator expansion.  
The complete set of operator identities for the ground state baryons
have been derived and
are given in the next section.  Construction of the independent operator basis 
${\cal O}_n$
of the baryon $1/N_c$ expansion for any spin $\times$ flavor representation
is straightforward using the operator identities.   

The formulation of the baryon $1/N_c$ expansion using static quark operators
is described in the work of Dashen {\it et al.}~\cite{dm,j,djm1,djm2}, 
Carone {\it et al.}~\cite{cgo} and Luty $\&$ March-Russell~\cite{lm,l}.
Dashen {\it et al.} showed that the $1/N_c$ expansion is given
in terms of operator products of the baryon spin-flavor generators, which 
can be interpreted as soliton operators or quark operators.
Carone {\it et al.} analyzed the $1/N_c$ expansion in terms of normal-ordered
static quark operators in the Hartree picture, while   
Luty $\&$ March-Russell studied the $1/N_c$
expansion in terms of static quark operators using a many-body approach.

\subsection{\it Redundant Operators (Operator Identities)}

The complete set of operator identities for
the ground state baryons have been derived by
Dashen {\it et al.}~\cite{djm2}.  The operator identities 
have an elegant group theoretic structure, and are given
in terms of the Casimir invariants of the $SU(2N_F)$ baryon spin-flavor
representation.  An important result of Ref.~\cite{djm2} is that
the only nontrivial operator
identities which are required are those which reduce $2$-body operators to
linear combinations of $1$-body or $0$-body operators.  All identities
for $n$-body operators with $n>2$ can be obtained by recursively applying the
$2$-body identities on all pairs of $1$-body operators appearing in a 
completely symmetric $n^{\rm th}$ order operator product.  
The operator identities have been derived for arbitrary numbers of 
light quark flavors $N_F$.

The operator identities for
$N_F=2$ light quark flavors are given in Table~1.  The spin $\otimes$ flavor
representation $(J,I)$ of each operator
identity is given explicitly.  The identities
are grouped according to their group theoretic origin.  The first identity
is the $SU(4)$ quadratic Casimir relation
\begin{equation}
 \Lambda^A \Lambda^A = C(R) \openone,
\end{equation}
where $\Lambda^A$, $A=1, \ldots, 15$, refer to the 15 $SU(4)$ generators
$J^i$, $I^a$ and $G^{ia}$, and $C(R)$ is the quadratic Casimir for the
baryon spin-flavor representation Fig.~6.
The second block of
identities results from the $SU(4)$ cubic Casimir relation 
\begin{equation}
d^{ABC}\Lambda^B \Lambda^C = D(R)\ \Lambda^A
\end{equation} 
which involves the $SU(4)$
$d$-symbol.  The third block of identities correspond to operators which
vanish identically when acting on the completely symmetric baryon spin-flavor 
representation.

The $1/N_c$ expansion for any static baryon operator can be 
constructed using the operator identities to reduce the set of operators
to a complete and independent set.  
It is possible to systematically construct an
operator basis which maximizes the number of $J$'s in operator products
by eliminating redundant operators which contain more powers of $I$ and $G$.
This reduction of operator products is summarized by an operator reduction
rule:  All operators in which 
two spin or isospin indices are contracted with a $\delta$ or $\epsilon$ symbol
can be eliminated, with the exception of $J^2$.  (Note $J^2 = I^2$ for the 
baryon states, so symmetry under spin and isospin remains.)
The application of the operator identities and the
operator reduction rule is straightforward,
and will be demonstrated in the following examples.    

\subsubsection{Masses}

The baryon mass transforms as a spin-isospin singlet $(0,0)$.  
The $0$-body operator $\openone$ transforms as a singlet, and is one of the
operators in the $1/N_c$ expansion.  
None of the baryon spin-flavor generators transforms as a singlet, so 
there are no $1$-body operators in the expansion.  
The candidate $2$-body operators
are the singlets $J^2$, $I^2$ and $G^2$.  The operator identities
give two nontrivial relations amongst these three $2$-body operators,
so only one of the three operators is independent.  Without loss of generality,
the $2$-body operator appearing in the $1/N_c$ expansion can be chosen to be
$J^2$.  This choice is in accord with the operator reduction rule.  
It is also
possible to classify the higher body operators 
using the $2$-body identities: the only higher body operators are $(J^2)^n$.     
Thus, the baryon mass has the expansion~\cite{j,cgo,lm}      
\begin{equation}\label{mexp1}
M = m_0 N_c \openone + m_2 {1 \over N_c} J^2 
+ m_4 {1 \over N_c^3} \left(J^2\right)^2\ldots ,
\end{equation}
where each term is multiplied by an arbitrary
coefficient, and the ellipsis represents higher body operators.
The mass expansion has the form
\begin{equation}\label{nm}
N_c {\cal P}\left( {1 \over N_c}, { J^2 \over {N_c^2}} \right),
\end{equation}
where ${\cal P}$ is a general polynomial in its arguments.
For $N_c =3$, the $1/N_c$ expansion
only extends to $3$-body operators, and there are only two terms in the mass
expansion Eq.~(\ref{mexp1}).  
The two mass coefficients $m_0$ and $m_2$ parametrize the two
masses $N$ and $\Delta$ in terms of the spin-independent average mass and the
hyperfine mass splitting $(\Delta -N)$.  For baryons with $J \sim O(1)$,
the hyperfine mass splitting is order $1/N_c$, and is suppressed by a factor of
$1/N_c^2$ relative to the leading $O(N_c)$ contribution to the baryon mass.
The mass splitting $(\Delta -N)$ also is proportional to $( <J^2>_{3/2} - 
<J^2>_{1/2}) = \frac {15} 4 -\frac 3 4 =3$, so one expects $(\Delta -N) \sim
300$~MeV given that the average $O(N_c)$ baryon mass is $\sim 1$~GeV.

\subsubsection{Axial Couplings}

The isovector baryon axial vector current transforms
as a $(1,1)$ under spin $\otimes$ isospin.  There is no $0$-body
operator transforming as a $(1,1)$.  
There is one $1$-body operator which
has spin-$1$ and isospin-$1$, namely $G^{ia}$.  
Two operator identities in Table 1 transform as a $(1,1)$.  
The $2$-body operators containing a single 
spin or flavor $\epsilon$-symbol in the second operator identity,
namely $\epsilon^{ijk}\{J^i, G^{jc} \}$ and $\epsilon^{abc}\{I^a, G^{kb} \}$,
do not transform under time-reversal
in the same way as the isovector axial
vector current, and therefore are not allowed operators.
The $2$-body operators in the first operator identity are both allowed
operators.  The two operators are not independent, so 
there
is one independent $2$-body operator, which can be chosen to be $J^i I^a$.
Again, this choice is prescribed by the operator reduction rule.
Higher-body operators are obtained by taking anticommutators of the above
operators with $J^2$.
The single allowed $3$-body operator is $\{ J^2, G^{ia} \}$, and the single
allowed $4$-body operator is $\{ J^2, J^i I^a \}$.  In general,   
the operators ${\cal O}_1^{ia} = G^{ia}$, ${\cal O}_2^{ia} = J^i T^a$ and 
\begin{equation}
{\cal O}^{ia}_{n+2} = \left\{ J^2, {\cal O}^{ia}_n \right\}
\end{equation}
for $n \ge 1$ form a complete and independent operator basis for the $1/N_c$
expansion of a $(1,1)$ operator.
Thus, the $1/N_c$ expansion for the isovector axial vector current is given by
\begin{equation}
A^{ia} = a G^{ia} + b {1 \over N_c} J^i I^a + c {1 \over N_c^2} 
\left\{ J^2, G^{ia} \right\}+ \ldots,
\end{equation}
where the ellipsis represents higher-body operators.  For $N_c=3$, there are
only three terms in expansion.  The three operator coefficients $a$, $b$ and
$c$ of the $1/N_c$ expansion parametrize the three isovector couplings 
$g_{\pi NN}$, $g_{\pi N \Delta}$ and $g_{\pi \Delta \Delta}$.
The matrix elements
of $G^{ia}$ are $O(N_c)$, and the matrix elements of $J$ and $I$ are $O(1)$
for low-spin baryons, so the $1/N_c$ expansion can be truncated after the
first term  
upto corrections of relative order $1/N_c^2$ from the neglected $b$ and $c$
terms.  Thus, one recovers
the result from Sect.~(2.4.2) that ratios of
the isovector
baryon axial vector couplings are given by $G^{ia}$ up to corrections of order
$1/N_c^2$ relative to the leading order.

\subsection{Hartree Picture}

The spin-flavor structure of large-$N_c$ baryons has been 
studied in the Hartree
picture by Carone {\it et al.}~\cite{cgo}.  A number of interesting results
can be derived by making the zeroth order ansatz that the Hartree potential is
spin and flavor independent.  

If the Hartree potential is spin and flavor
independent, then the $N_c$ quarks in the large-$N_c$ baryon are all in the
same ground state wavefunction $\phi(r)$, and the large-$N_c$ baryon
wavefunction is given by the product Eq.~(\ref{hartree}) of $N_c$ identical
quark wavefunctions.   
The baryon wavefunction must be completely
symmetric in space, spin and flavor, so the complete symmetry of the baryon
wavefunction with respect to the spatial coordinates of the $N_c$ valence 
quarks implies that the baryon wavefunction is completely symmetric 
in the spin-flavor indices of the $N_c$ quarks.  Thus,  
the spin-flavor wavefunction of the ground state baryons
is given by the completely 
symmetric spin-flavor representation shown in Fig.~6.

A generic $n$-body operator ${\cal O}_n$ appearing in the $1/N_c$ expansion 
for baryons can be written as
\begin{eqnarray}\label{hartreeop}
&&\int d^3 {\bf x}_{1} \ldots d^3 {\bf x}_{n}
\phi_{\jmath_1\beta_1}^\dagger\left({\bf x}_{1}\right)\ldots
\phi_{\jmath_n\beta_n}^\dagger\left({\bf x}_{n}\right)\times \nonumber\\ 
&&\qquad\qquad{\cal O}_{(\imath_1\alpha_1) 
\ldots (\imath_n\alpha_n)}^{(\jmath_1\beta_1) \ldots (\jmath_n\beta_n)}
\left({\bf x}_{1},\ldots,{\bf x}_{n}\right)
\phi^{\imath_1\alpha_1}\left({\bf x}_{1}\right)\ldots
\phi^{\imath_n\alpha_n}\left({\bf x}_{n}\right) \ ,
\end{eqnarray}
where $\phi^{\imath\alpha}\left({\bf x}\right)$ denotes a quark annihilation
operator which 
annihilates a quark with
spin $\imath$ and flavor $\alpha$ in a given wavefunction.
The ansatz for the ground state baryon wavefunction implies that the quark 
wavefunction annihilation operator can be replaced by 
\begin{equation}
\phi^{\imath\alpha}\left({\bf x}\right) \rightarrow \phi(r) q^{\imath\alpha},
\end{equation}
where the quark annihilation operator $q^{\imath\alpha}$ only acts on the spin
and flavor indices of the quark.
The spatial symmetry of the baryon wavefunction implies that 
the tensor $\cal O$ in Eq.~(\ref{hartreeop}) does not carry 
any angular momentum, because terms transforming nontrivially 
under orbital angular momentum on any quark line will vanish when 
integrated over space.  Thus, the operator tensor may be replaced by
\begin{equation}
{\cal O}_{(\imath_1\alpha_1) 
\ldots (\imath_n\alpha_n)}^{(\jmath_1\beta_1) \ldots (\jmath_n\beta_n)}
\left({\bf x}_{1},\ldots,{\bf x}_{n}\right)
\rightarrow{\cal O}_{(\imath_1\alpha_1) 
\ldots (\imath_n\alpha_n)}^{(\jmath_1\beta_1) \ldots (\jmath_n\beta_n)}
\left({r}_{1},\ldots,{r}_{n}\right)
\end{equation}
when acting on the ground state baryons.  After the spatial integrations in
Eq.~(\ref{hartreeop}) are performed, the Hartree operator ${\cal O}_n$
reduces to a normal-ordered quark operator acting on the spin and 
flavor indices of $n$ static quarks,   
Eq.~(\ref{normalordered}).  

The above deductions are predicated on the assumption that the 
Hartree potential is spin-independent at leading order in $1/N_c$.  
The Hartree Hamiltonian is a spin and flavor singlet, and has an 
expansion of the form
\begin{equation}
{\cal H} = h_0 N_c \openone + h_2 {1 \over N_c} J^2 + \ldots,
\end{equation} 
where $h_n$ are arbitrary unknown coefficients.  
For baryons with spins $J \sim O(1)$, the spin-dependent part of the Hartree
potential is suppressed in $1/N_c$ relative to the spin-independent part, and
the initial assumption that the Hartree potential is spin and flavor
independent is justified in the large-$N_c$ limit.  The spin-dependent part of
the Hartree potential first appears at relative order $1/N_c^2$ compared to
the leading term, so the wavefunction of low-spin baryons is not
significantly deformed by spin-dependent interactions.   
Thus, the Hartree picture
leads to a self-consistent picture of the spin-flavor structure of large-$N_c$
baryons with spins of order unity. 
For baryons with spins $J \sim O(N_c)$, the spin-dependent contributions to 
the energy are not corrections, but are $O(N_c)$ and of leading order, so  
the ansatz that the Hartree potential is spin and flavor 
independent is not valid, and the argument breaks down.  
Significant deformation of the baryon wavefunction due to spin-dependent
interactions is possible for baryons with spins $J \sim O(N_c)$.

\subsection{\it $SU(3)$ Flavor Symmetry}

The baryon $1/N_c$
expansion for $N_F=3$ light quark flavors is considerably more complicated 
than for two light flavors, because baryon states and operator matrix elements
have a nontrivial dependence on $N_c$.

The ground state baryon representation of the
$SU(2N_F)$ spin-flavor algebra is given by
the completely symmetric Young tableau with $N_c$ boxes.  For odd $N_c$,
the decomposition of this spin-flavor representation under 
$SU(2N_F) \rightarrow SU(2) \otimes SU(N_F)$ yields the baryon spin tower 
of Fig.~7.
For any $N_F \ge 3$,   
the flavor representations of the baryon spin tower have dimensions which
depend on $N_c$ even for low-spin baryons.
The $SU(3)$ flavor weight diagrams for the spin $\frac 1 2$ and spin
$\frac 3 2$ baryons are shown explicitly in Figs.~10 and~11,
respectively.
The weight diagrams reduce to octet and decuplet multiplets for $N_c=3$,
but there are many additional states for $N_c > 3$.  
Because the baryon flavor representations vary with $N_c$, 
the identification of QCD baryons with large-$N_c$ baryons is not unique.    
The $1/N_c$ expansion
yields the most predictions for baryons with finite spin and flavor quantum
numbers, so it is best to identify the QCD baryons with states at the top of
the weight diagrams with fixed finite strangeness.  
Note that the large-$N_c$ baryon states in the $SU(3)$ flavor weight diagrams
with a given fixed strangeness
are in one-to-one correspondence 
with the $(J,I)$ towers of Sect. (2.3).  
Although the identification of QCD baryons with large-$N_c$ baryons is not
unique, the extrapolation of the $1/N_c$ expansion of baryon operators to
$N_c=3$ is unambiguous, and all the $1/N_c$ expansions in the literature
reduce to the same operator expansion for $N_c =3$.

The analysis of the $N_c$-dependence of operator products appearing in the
$1/N_c$ expansion is subtle for $N_F=3$ light quark flavors because the
operator matrix elements have a different $N_c$-dependence in different parts
of the flavor weight diagram.  The $N_c$-dependence of operator products 
is determined by the $N_c$-dependence of the baryon spin-flavor generators.
The $SU(6)$ spin-flavor generators are given by 
\begin{eqnarray}
&&J^i = q^\dagger \left({ \sigma^i \over 2} \otimes \openone \right) q,
\nonumber \\
&&T^a = q^\dagger \left(\openone \otimes {\lambda^a \over 2} \right) q,\\
&&G^{ia} = q^\dagger \left({\sigma^i \over 2} 
\otimes {\lambda^a \over 2}\right) q, \nonumber
\end{eqnarray}
where $\lambda^a$ are Gell-Mann $SU(3)$ matrices.
The matrix elements of the baryon spin $J$ are $O(1)$ for baryons at the bottom
of the spin tower and $O(N_c)$ for baryons at the top of the spin tower, as
before.
The matrix elements of the other spin-flavor generators are
much more subtle. 
Unlike the
case for two light flavors, matrix elements of the flavor generators $T^a$
and the matrix elements of the spin-flavor generators
$G^{ia}$
do not have the same $N_c$-dependence everywhere in the flavor weight diagram.
Consider baryons at the top of the flavor weight diagrams 
with finite strangeness.  These baryons have matrix
elements of the flavor generators $T^a$, which are $O(1)$, $O(\sqrt{N_c})$ and
$O(N_c)$ for $a=1,2,3$, $a=4,5,6,7$, and $a=8$, respectively, and matrix
elements of the spin-flavor generators $G^{ia}$ which are $O(N_c)$,
$O(\sqrt{N_c})$ and $O(1)$, for $a=1,2,3$, $a=4,5,6,7$, and $a=8$, respectively.
For baryon states in other regions of the weight diagrams, 
different combinations of the $T$'s
and $G$'s will be order $N_c$, $\sqrt{N_c}$ and $1$.  
Thus, the matrix elements of $T^a$ are not suppressed relative to matrix
elements of $G^{ia}$ everywhere in the flavor weight diagram, and 
it is not possible to neglect the matrix elements of $T^a$ or $G^{ia}$ 
for any $a$ unless one restricts one's attention to baryons in a certain
region of the flavor weight diagram, such as baryons with finite flavor
quantum numbers.  Irregardless of this choice, the baryon $1/N_c$ expansion
is predictive for baryons with $J \sim O(1)$ because the matrix
elements of the operator products with more powers of $J/N_c$ are suppressed.

The operator identities for $N_F=3$ light quark flavors are given in
Table~2, which is taken from Ref.~\cite{djm2}.
Many new operator structures appear at $N_F=3$ because there is
a $d$-symbol for the $SU(3)$ flavor group.
The elimination of redundant operators is not as simple for three light
quark flavors as for two light quark flavors, so the operator reduction rule
is modified.       
The operator reduction rule for $N_F=3$ light quark flavors is:
All operator products in which two flavor indices are contracted using
$\delta^{ab}$, $d^{abc}$, or $f^{abc}$ or two spin indices on $G$'s are
contracted using $\delta^{ij}$ or $\epsilon^{ijk}$ can be eliminated.

\subsection{\it Flavor Symmetry Breaking}

An additional complication for $N_F=3$ light quark flavors is that effects of
$SU(3)$ flavor symmetry breaking are comparable to $1/N_c$
corrections, so flavor symmetry breaking cannot be neglected.
$SU(3)$ flavor symmetry breaking can be included perturbatively in the $1/N_c$
expansion.  The combined expansion in $1/N_c$ and $SU(3)$ flavor symmetry
breaking produces a rich pattern of spin-flavor symmetry breaking which is
not dominated either by $1/N_c$ or flavor symmetry breaking corrections. 

The baryon $1/N_c$ expansion for finite $N_c$ only extends to operator products
of $N_c^{\rm th}$ order in the baryon spin-flavor generators, so the
perturbative expansion in $SU(3)$ flavor breaking extends only to a given
finite order in flavor symmetry breaking.  For a flavor and spin singlet
baryon operator with a $1/N_c$ expansion beginning with the $0$-body operator
$\openone$, the expansion in flavor symmetry breaking goes to $N_c^{\rm th}$
order.  For baryon operators with $1/N_c$ expansions beginning with a $1$-body
operator, such as a flavor octet or a spin-$1$ baryon operator, the flavor
symmetry breaking expansion goes to order $(N_c -1)$.  In general, the flavor
symmetry breaking expansion extends to order $(N_c -n)$ for a baryon operator 
with a $1/N_c$ expansion beginning with a $n$-body operator.  

The incorporation of flavor symmetry breaking into the baryon $1/N_c$ expansion
is best illustrated by an example.

\subsubsection{$SU(3)$ Flavor Analysis of Baryon Masses}

The masses of the baryon octet and decuplet can be analyzed in a combined
expansion in $1/N_c$ and $SU(3)$ flavor symmetry breaking~\cite{jl}.  
The combined expansion is given by
\begin{equation}
M = M^1 + M^8 + M^{27} + M^{64},
\end{equation}
where the $1/N_c$ expansions of the
flavor singlet, octet, $\bf 27$ and $\bf 64$ contributions to the baryon masses
are 
\begin{eqnarray}\label{su3masses}
&&M^1 = N_c \openone + {1 \over N_c}J^2, \nonumber\\
&&M^8 = T^8 + {1 \over N_c} \left\{J^i, G^{i8} \right\}
+ {1 \over N_c^2} \left\{ J^2, T^8 \right\}, \nonumber\\
&&M^{27} = {1 \over N_c}\left\{ T^8, T^8 \right\} 
+{1 \over N_c^2}\left\{ T^8, \left\{ J^i, G^{i8} \right\}\right\}, \\
&&M^{64} = {1 \over N_c^2}\left\{ T^8, \left\{ T^8, T^8
\right\}\right\},\nonumber
\end{eqnarray}
respectively, and it is to be understood that each operator is multiplied by
an unknown coefficient.  
The $1/N_c$ expansion extends only up to $3$-body operators for QCD baryons with
$N_c=3$.  The coefficients of the flavor singlet operators in $M^1$ are $O(1)$
in $SU(3)$ flavor symmetry breaking, while the coefficients of the flavor-octet
operators in $M^8$ are proportional to one power of $SU(3)$ flavor symmetry
breaking $\epsilon$.  The $SU(3)$ flavor breaking parameter $\epsilon$ is
proportional to $m_s/\Lambda_\chi$, and is order $30\%$.    
The flavor-$\bf 27$ operators are second order in $SU(3)$
flavor symmetry breaking, and the one flavor-$\bf 64$ operator is third order in
$SU(3)$ flavor symmetry breaking.  Each operator also occurs at a definite order
in $1/N_c$, which is given by the explicit factor of $1/N_c$ in front of the
operator times the leading $N_c$-dependence of the operator matrix element.   

The eight mass operators in the $1/N_c$
expansion Eq.~(\ref{su3masses}) parametrize the eight baryon masses $N$,
$\Delta$, $\Lambda$, $\Sigma$, $\Sigma^*$, $\Xi$, $\Xi^*$, and $\Omega$ of the
spin-$\frac 1 2$ octet and spin-$\frac 3 2$ decuplet baryons.  Each operator
of the combined $1/N_c$ and flavor symmetry breaking expansion corresponds to 
a unique linear combination of octet and decuplet baryon masses.
Mass relations between octet and decuplet baryon masses are obtained by
neglecting subleading operators in the combined $1/N_c$ and flavor symmetry 
breaking expansion.  
For example, the mass combination corresponding to the $J^2/N_c$ operator is 
\begin{equation}\label{ocdechyper}
\frac 1 8 \left( 2 N + \Lambda + 3 \Sigma + 2 \Xi \right) - \frac 1 {10}
\left( 4 \Delta + 3 \Sigma^* + 2 \Xi^* + \Omega \right) .
\end{equation}  
Neglect of the $J^2/N_c$ operator relative to the leading singlet
operator implies that the mass combination Eq.~(\ref{ocdechyper}) vanishes.
Thus, the average mass of the baryon octet is equal to the average mass
of the average mass of the baryon decuplet when the hyperfine operator $J^2/N_c$
is neglected.  The eight baryon mass combinations
baryon mass combinations are graphed in Fig.~12.
The dimensionless
quantity $\sum B_i / (\sum |B_i|/2)$ is graphed for each mass combination.
The hierarchy of baryon masses is in excellent agreement with the $1/N_c$ and
flavor symmetry breaking factors of the combined analysis. 

\section{SUMMARY OF MAIN RESULTS}

Many results have been derived in the literature using the baryon $1/N_c$
expansion.  A brief description of the some of the main applications is
given here.

\subsection{\it Axial Couplings}

The baryon axial vector couplings have been studied extensively.  The
analysis for $N_F=2$ light quark flavors is given in the original paper
of Dashen $\&$ Manohar~\cite{dm}.
The extension to $N_F=3$ light quark
flavors with arbitrary $SU(3)$ flavor symmetry breaking is given by
Dashen {\it et al.}~\cite{djm1} using the operator $X_0^{ia}$.  
The analysis with exact $SU(3)$ flavor symmetry imposed 
at leading order in $1/N_c$ also is given.  
In Luty~\cite{l}, the $SU(3)$ flavor symmetry analysis at leading order
in $1/N_c$ is given using quark operators.  
The complete $1/N_c$ expansion 
with exact $SU(3)$ flavor symmetry and perturbative $SU(3)$ breaking is derived
in Dashen {\it et al.}~\cite{djm2}.  A detailed fit to the hyperon decay data
and measured decuplet $\rightarrow$ octet + pion axial couplings is performed
in Ref.~\cite{ddjm}.   

The two-flavor analysis was presented in Sect.~2 of this review.  For
three light quark flavors, a number of additional results are found.
In the $SU(3)$ flavor symmetry limit, the pion couplings of the spin-$\frac 1 2$
octet and spin-$\frac 3 2$ decuplet baryons are described by the $1/N_c$
expansion~\cite{djm1,l,djm2}
\begin{eqnarray}\label{su3pion}
A^{ia} &=& a_1 G^{ia} + b_2 {1 \over N_c } J^i T^a + b_3 {1 \over N_c^2}
\left\{J^i,\left\{J^j, G^{ja} \right\} \right\} \\
&&+ c_3 {1 \over N_c^2} \left(\left\{ J^2, G^{ia} \right\} - \frac 1 2 
\left\{J^i,\left\{J^j, G^{ja} \right\} \right\} \right). \nonumber 
\end{eqnarray}
The four coefficients $a_1$, $b_2$, $b_3$ and $c_3$ parametrize the four
pion couplings $D$ and $F$ of the baryon octet, ${\cal C}$ of
decuplet-to-octet transitions, and ${\cal H}$ of the decuplet~\cite{bchpt}.  
At leading order in $1/N_c$, the expansion can be
truncated after two operators~\cite{djm1,l,djm2}, and one obtains the two
relations
\begin{equation}
{\cal C} =-2D, \qquad {\cal H}=3D-9F,
\end{equation}
which are valid upto corrections of relative order $1/N_c^2$.  The first relation
is an $SU(6)$ relation, which explains why an $SU(6)$ coupling is found
in chiral perturbation theory calculations~\cite{bchpt}.

The $F/D$ ratio can be extracted by analyzing the baryon-pion couplings for
arbitrary large $N_c$~\cite{djm1,djm2}.  
If QCD baryons are identified with 
the large-$N_c$ baryons at the top of the
$SU(3)$ flavor weight diagrams in Figs.~10 and~11
with strangeness of order unity, then the $b_2$ term
in Eq.~(\ref{su3pion}) is of relative order $1/N_c^2$ compared
to the $a_1$ term for the pion couplings $a=1,2,3$; 
of relative order $1/N_c$ for the kaon couplings
$a=4,5,6,7$; and of relative order $1$ for the $\eta$ couplings $a=8$.
Using the pion couplings of baryons with finite strangeness, one obtains
$F/D = 2/3 +O(1/N_c^2)$.  The determinations from the kaon and $\eta$ couplings
are $F/D = 2/3 +O(1/N_c)$ and $F/D = 2/3 +O(1)$, respectively.

$SU(3)$ symmetry breaking leads to a rich hierarchy of
couplings~\cite{djm2,ddjm}.
One particularly interesting new result,
mentioned in Sect.~(2.4.2),
is that
there is an equal spacing rule for the pion axial vector couplings
to baryons with differing strangeness~\cite{djm1}.  This feature of the
symmetry breaking is seen in the experimental data.

\subsection{\it Masses}

The baryon masses have been analyzed in many contexts.
The first
derivation of the subleading $J^2/N_c$ hyperfine operator is given
by Jenkins~\cite{j}.  The three-flavor analysis with $SU(3)$ flavor breaking
is given by Dashen {\it et al.}~\cite{djm1}.  The analysis was performed
using the operator $\bf K$.
The same mass relations subsequently were obtained by Luty~\cite{l} using
quark operators with manifest $SU(3)$ flavor symmetry.  
The large-$N_c$ result Eq.~(\ref{nm}) for
exact flavor symmetry is given by Carone {\it et al}~\cite{cgo} and Luty $\&$
March-Russell~\cite{lm}.
Jenkins $\&$ Lebed~\cite{jl} analyzed the isospin $I=0,1,2,3$ mass splittings
of the baryon octet and decuplet to all orders in 
a combined expansion in $1/N_c$ and flavor
symmetry breaking.  The analysis from Ref.~\cite{jl}
of the $I=0$ isospin-averaged masses
was presented in Sect.~(3.5.1).  Results
also were obtained for the isospin splittings of the baryon masses.  A sample
result of this analysis is that the Coleman-Glashow mass relation,
\begin{equation}
\left(p-n\right) - \left(\Sigma^+ - \Sigma^-\right)+\left(\Xi^0 - \Xi^-\right)=0,
\end{equation}   
is $O(1/N_c)$ in the $1/N_c$ expansion, so the mass relation should be more
accurate than is predicted by a flavor symmetry breaking analysis alone.
Many additional predictions for the isospin splittings are found as well.

Nonanalytic corrections to the baryon masses have been calculated in chiral
perturbation theory in conjunction with the $1/N_c$ expansion in
Refs.~\cite{j,djm1,lm,jchpt,bedaque}.  The leading $m_s^{3/2}$ correction to the
flavor-$\bf 27$ baryon masses of the octet and decuplet is calculated in
Jenkins~\cite{jchpt}.  Subleading chiral corrections are considered in Bedaque
$\&$ Luty~\cite{bedaque}.   

\subsection{\it Magnetic Moments}

The baryon magnetic moments have been studied by many authors.  The first result
by Dashen $\&$ Manohar~\cite{dm} shows that the isovector baryon magnetic moments
are proportional to $X_0^{ia}$ in the large-$N_c$ limit upto a correction of
relative order $1/N_c^2$, so the ratios of the isovector
magnetic moments are determined for $N_F=2$ flavors
upto a correction of relative order $1/N_c^2$.
The leading $m_s^{1/2}$ chiral
correction to the baryon magnetic moments is observed to be $O(N_c)$ by Luty {\it
et al}~\cite{lmw}.  A flavor $SU(3)$ analysis of the isoscalar and isovector
magnetic moments including $SU(3)$ flavor symmetry breaking is given by Jenkins
$\&$ Manohar~\cite{jm}.  
An $SU(3)$ group theory decomposition of the leading
chiral correction is given in Ref.~\cite{ddjm}, and a detailed comparison is made
to the experimental data.

The magnetic moments are found to satisfy the $1/N_c$ hierarchy predicted by the
$1/N_c$ expansion.  The accuracy of magnetic moment relations is in good
agreement with the theoretical predictions.

\subsection{\it Other Applications}

Many additional applications of the $1/N_c$ expansion have been considered in the
literature.  A few of the more important applications are briefly remarked upon
below.

\subsubsection{Heavy Quark Baryons}

The couplings of baryons containing a single heavy quark to pions and mesons
containing a single heavy quark are considered in
Ref.~\cite{j}.  The pion couplings of the heavy quark baryons are related
to the pion couplings of the baryon octet and decuplet.

The masses of heavy baryons containing a single charm or bottom quark are
studied in Ref.~\cite{jhqet}.  The heavy baryons
are shown to satisfy a heavy quark spin-flavor symmetry as well as a large-$N_c$
light quark spin-flavor symmetry.  The masses are studied to all orders in a
combined expansion $1/N_c$, $SU(3)$ flavor symmetry breaking $\epsilon$, 
and the heavy quark
symmetry breaking parameter $\sim\Lambda_{\rm QCD}/m_Q$.   
Mass relations between charm baryon splittings
and bottom baryon splittings are obtained, as are relations between mass
splittings of heavy quark baryons and the baryon octet and decuplet.  The most
accurate mass relations can be used to predict the masses of unmeasured charm and
bottom baryons.

\subsubsection{Excited Baryons}

Excited baryons also can be studied in the $1/N_c$ expansion.  For finite large
$N_c$, the first 
excited baryons are identified with the $SU(2N_F)$ representation
shown in Fig.~13.
The pion couplings of the excited baryons to the
ground state baryons are studied in the large-$N_c$ limit in Carone {\it et
al.}~\cite{cgkm} using quark operators.  Pirjol $\&$ Yan~\cite{py} 
derived large-$N_c$
consistency conditions for the pion couplings of the excited baryons and analyzed
the masses and mixings of the baryons.  Many of the large-$N_c$
predictions for excited
baryons are seen in the observational data~\cite{py2}.   

\subsubsection{Nuclear Physics}

The nuclear potential has been studied in the $1/N_c$
expansion~\cite{kaplansavage,kaplanmanohar}.  The $1/N_c$ suppressions of 
various phenomenological terms can be predicted, and are found to accurately
reproduce the known magnitudes of these terms.

\subsubsection{Chiral Perturbation Theory}

Large-$N_c$ consistency conditions can be derived by studying chiral loop
corrections to baryon amplitudes.  The constraint that the chiral loop
corrections do not grow with more powers of $N_c$ than the leading order
tree-level terms leads to consistency conditions equivalent to the ones
presented in this review~\cite{dm,j}.  Cancellations between loop
corrections with different intermediate baryons
are required at leading order in $1/N_c$
for consistency of the $N_c$ counting of the chiral
expansion~\cite{dm,j,djm1}.

In Ref.~\cite{jchpt}, the baryon chiral Lagrangian is formulated in an expansion
in $1/N_c$ for finite $N_c$ to all orders in $1/N_c$, and 
it is shown how to how to perform chiral loop calculations at finite $N_c$.

\section{CONCLUSIONS}

The $1/N_c$ expansion has many important implications for the spin and flavor
properties of baryons.  Large-$N_c$ baryons satisfy a contracted spin-flavor
algebra, which is used to organize the group theory of the baryon $1/N_c$
expansion.  The baryon $1/N_c$ expansion is formulated in terms of operator
products of the baryon spin-flavor generators, whose baryon matrix elements are
known.  The spin-flavor structure and $1/N_c$ suppression factors of the baryon
operator expansion are determined.
A wide variety of symmetry
predictions and relations for baryon static matrix elements can be obtained by
neglecting subleading operators in the baryon $1/N_c$ expansion.  
The accuracy of
various symmetry relations is predicted by the $1/N_c$ expansion.    
The spin-flavor algebra of large-$N_c$ baryons connects the quark and soliton
pictures of large-$N_c$ baryons.  Many successes of the nonrelativistic quark
model and the Skyrme model can be understood in terms of the spin-flavor
structure of the baryon $1/N_c$ expansion in large-$N_c$ QCD.

\subsubsection*{ACKNOWLEDGMENTS}

This work was supported in part by a fellowship from the Alfred P. Sloan
Foundation, National Young Investigator Award PHY-9457911
from the National Science Foundation, and grant DOE-FG03-97ER40546
from the Department of Energy.

\begin{table}[htbp]
\caption{$SU(4)$ Operator Identities for $N_F =2$.}
\smallskip
\label{tab:su4iden}
\centerline{\vbox{ \tabskip=0pt \offinterlineskip
\def\tablerule{\noalign{\hrule}}
\def\space{height 4pt&\omit&&\omit&\cr}
\halign{
\vrule #&\strut\hfil\ $ # $\ \hfil&\vrule #&\strut\hfil\ $ # $\ 
\hfil&\vrule #\cr
\tablerule\space
&\left\{J^i,J^i\right\} + \left\{I^a,I^a\right\} +
4\ \left\{G^{ia},G^{ia}\right\} ={3 \over2} N \left(N+4\right)
&&(0,0)&\cr
\space\tablerule\space
&2 \left\{J^i,G^{ia}\right\} = \left(N+2\right)\ I^a
&&(0,1)&\cr
\space
&2 \left\{I^a,G^{ia}\right\} = \left(N+2\right)\ J^i
&&(1,0)&\cr
\space
&{1\over 2}\ \left\{J^k,I^c\right\}
-\epsilon^{ijk} \epsilon^{abc} \left\{G^{ia}, G^{jb}\right\} = 
\left(N+2\right)\ G^{kc} 
&&(1,1)&\cr
\space\tablerule\space
&\left\{I^a,
I^a\right\} - \left\{J^i,J^i\right\}=0 &&(0,0)&\cr
\space
&4\ \left\{G^{ia},G^{ib}\right\} = \left\{I^a,I^b\right\}\qquad (I=2)
&&(0,2)&\cr
\space
&\epsilon^{ijk}\ \left\{ J^i,G^{jc}\right\} = \epsilon^{abc} \ 
\left\{I^a,G^{kb} 
\right\}&&(1,1)&\cr
\space
&4\ \left\{G^{ia}, G^{ja}\right\} = \left\{J^i, J^j
\right\}\qquad (J=2)&&(2,0)&\cr
\space\tablerule
}}}
\end{table}

\begin{table}[htbp]
\caption{$SU(6)$ Operator Identities for $N_F=3$.}
\smallskip
\label{tab:su6iden}
\centerline{\vbox{ \tabskip=0pt \offinterlineskip
\def\tablerule{\noalign{\hrule}}
\def\space{height 4pt&\omit&&\omit&\cr}
\halign{
\vrule #&\strut\hfil\ $ # $\ \hfil&\vrule #&\strut\hfil\ $ # $\ 
\hfil&\vrule #\cr
\tablerule\space
&2\ \left\{J^i,J^i\right\} + 3\ \left\{T^a,T^a\right\} + 12\ 
\left\{G^{ia},G^{ia}\right\} = 5 N \left(N+6\right)&&(0,0)&\cr
\space\tablerule\space
&d^{abc}\ \left\{G^{ia}, G^{ib}\right\} + {2\over 3}\ \left\{J^i,G^{ic}
\right\} + {1\over4}\ d^{abc}\ \left\{T^a, T^b\right\} = {2\over 3}
\left(N+3\right)\ T^c  && (0,8) &\cr
\space
&\left\{T^a,G^{ia}\right\} = {2\over3}\left(N+3\right)\ J^i && (1,0)&\cr
\space&{1\over 3}\ \left\{J^k,T^c\right\} +  d^{abc}\ \left\{T^a,G^{kb}\right\}
-\epsilon^{ijk} f^{abc} \left\{G^{ia}, G^{jb}\right\} =  {4\over3}
\left(N+3\right)\ G^{kc} && (1,8)&\cr
\space\tablerule\space
&-12\ \left\{G^{ia},G^{ia}\right\} + 27\ \left\{T^a,
T^a\right\} - 32\ \left\{J^i,J^i\right\}=0 && (0,0)&\cr
\space
&d^{abc}\ \left\{G^{ia}, G^{ib}\right\} + {9\over 4} \ d^{abc}\ \left\{
T^a, T^b\right\} - {10\over3}\ \left\{J^i,G^{ic}\right\} = 0 && (0,8)&\cr
\space
&4\ \left\{G^{ia},G^{ib}\right\} = \left\{T^a,T^b\right\}\qquad ({27})
&& (0,{27})&\cr
\space
&\epsilon^{ijk}\ \left\{ J^i,G^{jc}\right\} = f^{abc} \ \left\{T^a,G^{kb}
\right\}&& (1,8)&\cr
\space
&3\ d^{abc}\ \left\{T^a,G^{kb}\right\} = \left\{J^k,T^c\right\} -  
\epsilon^{ijk} f^{abc}\ \left\{G^{ia}, G^{jb}\right\}&& (1,8)&\cr
\space
&\epsilon^{ijk}\ \left\{G^{ia},G^{jb}\right\} = f^{acg} d^{bch}\ \left\{
T^g,G^{kh}\right\}\qquad ({10}+{\overline {10}}) && 
(1,{10}+{\overline {10}})&\cr
\space
&3\ \left\{G^{ia}, G^{ja}\right\} = \left\{J^i, J^j
\right\}\qquad (J=2) && (2,0)&\cr
\space
&3\ d^{abc}\ \left\{G^{ia}, G^{jb}\right\} = 
\left\{J^i,G^{jc}\right\}\qquad (J=2) && (2,8)&\cr
\space\tablerule
}}}
\end{table}

\vfil\break\eject

\def\ssqr#1#2{{\vbox{\hrule height #2pt
      \hbox{\vrule width #2pt height#1pt \kern#1pt\vrule width #2pt}
      \hrule height #2pt}\kern- #2pt}}
\def\sqr{\mathchoice\ssqr8{.4}\ssqr8{.4}\ssqr{5}{.3}\ssqr{4}{.3}}
\def\bsqr{\ssqr{10}{.2}}
\def\nbox{\vbox{\hbox{$\bsqr\bsqr\bsqr\bsqr\raise2.7pt\hbox{$\,\cdot\cdot
\cdot\cdot\cdot\,$}\bsqr\bsqr\bsqr$}\nointerlineskip
\kern-.2pt\hbox{$\phantom{\bsqr}$}}}

\def\ndots{\vbox{\hbox{$\phantom{\bsqr}\raise2.7pt\hbox{$\,\cdot\cdot\cdot
\,$}$}\nointerlineskip
\kern-.2pt\hbox{$\phantom{\bsqr}$}}}

\def\nboxA{\vbox{\hbox{$\bsqr\bsqr\bsqr\bsqr\raise2.7pt\hbox{$\,\cdot\cdot
\cdot\cdot\cdot\,$}\bsqr\bsqr\bsqr$}\nointerlineskip 
\kern-.2pt\hbox{$\bsqr$}}}

\def\nboxE{\vbox{\hbox{$\bsqr\bsqr\bsqr\raise2.7pt\hbox{$\,\cdot\cdot
\cdot\cdot\cdot\,$}\bsqr\bsqr\bsqr\bsqr$}\nointerlineskip 
\kern-.2pt\hbox{$\bsqr\bsqr\bsqr\raise2.7pt\hbox{$\,\cdot\cdot\cdot
\cdot\cdot\,$}\bsqr$}}}

\def\nboxF{\vbox{\hbox{$\bsqr\bsqr\bsqr\bsqr\raise2.7pt\hbox{$\,\cdot
\cdot\cdot\cdot\cdot\,$}\bsqr\bsqr$}\nointerlineskip 
\kern-.2pt\hbox{$\bsqr\bsqr\bsqr\bsqr\raise2.7pt\hbox{$\,\cdot\cdot
\cdot\cdot\cdot\,$}\bsqr$}}}

\begin{figure}\label{doubleline}
\centerline{\epsfxsize=1.5in\epsffile{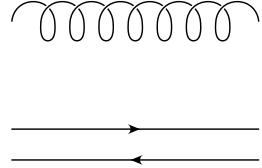}}
\caption{Double line notation for a gluon.}
\end{figure}

\begin{figure}\label{threemeson}
\centerline{\epsfxsize=1.5in\epsffile{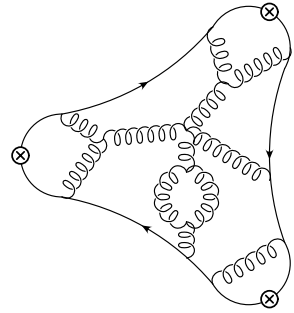}}
\caption{A typical planar diagram contributing to a meson three-point vertex
at leading order.}
\end{figure}

\begin{figure}\label{bmbm}
\centerline{\epsfxsize=1.5in\epsffile{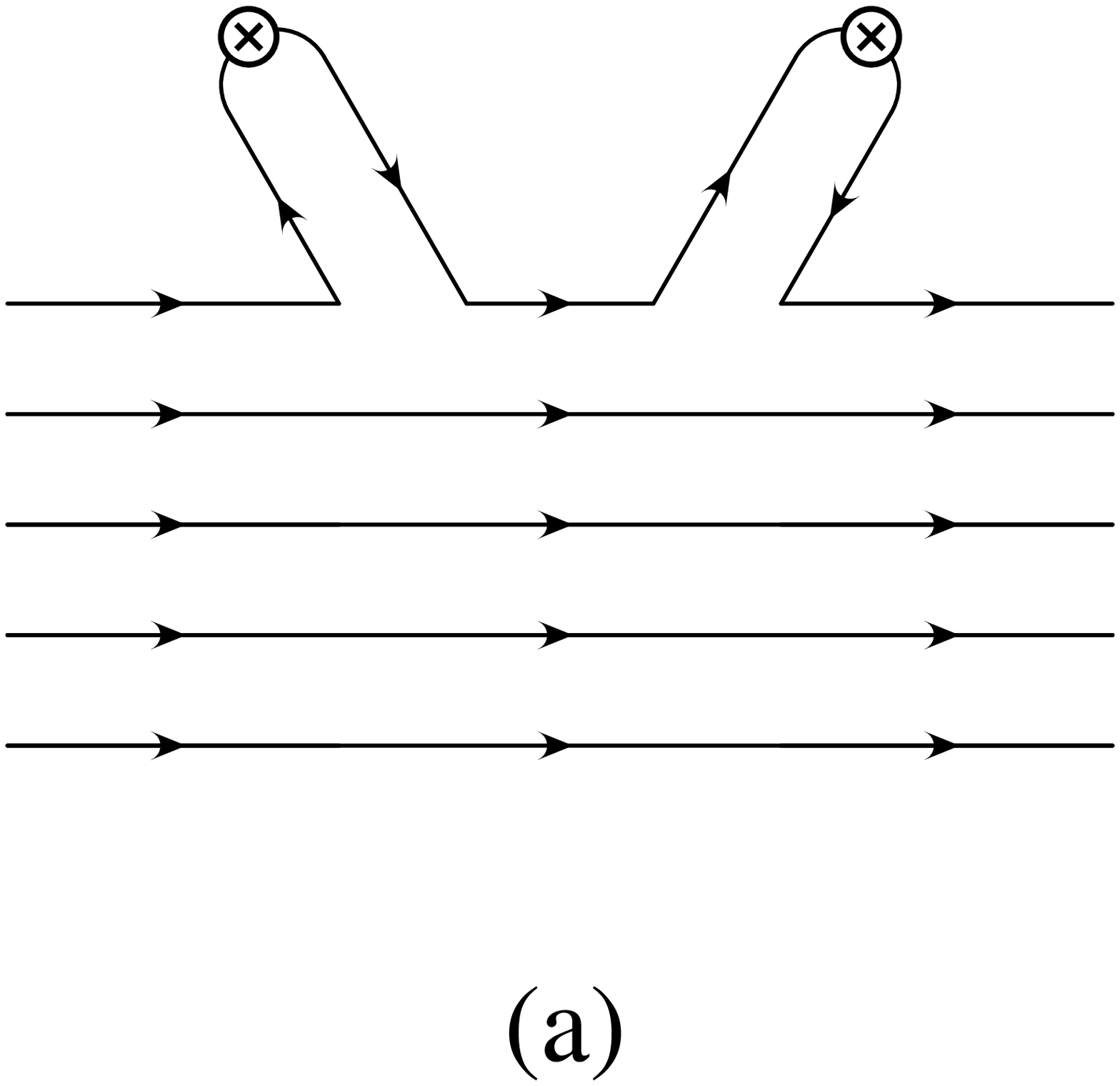}\qquad
\epsfxsize=1.5in\epsffile{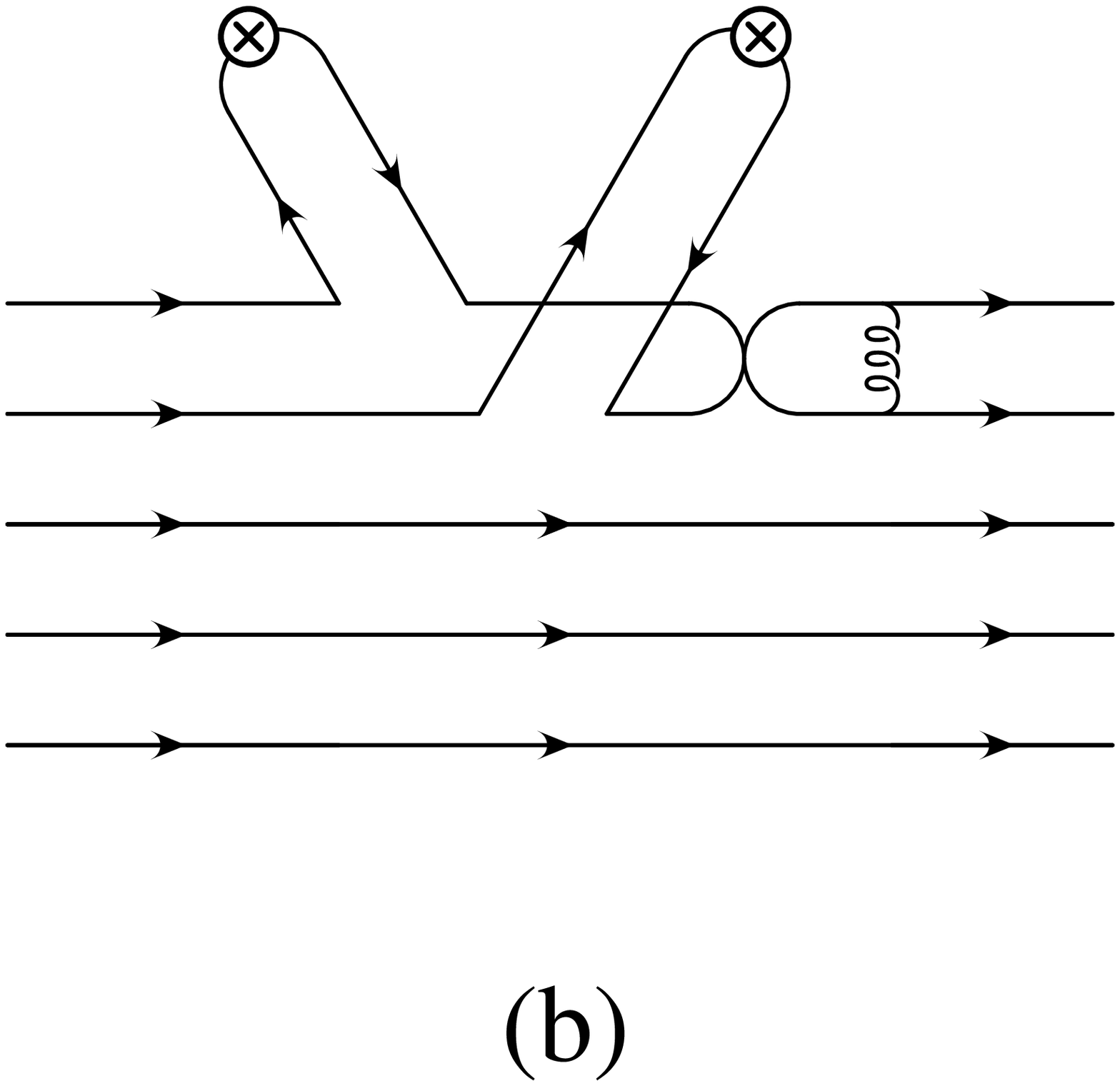}}
\caption{Quark-gluon diagrams for the scattering process baryon + meson
$\rightarrow$ baryon + meson.}
\end{figure}

\begin{figure}\label{bmvertex}
\centerline{\epsfxsize=1.2in\epsffile{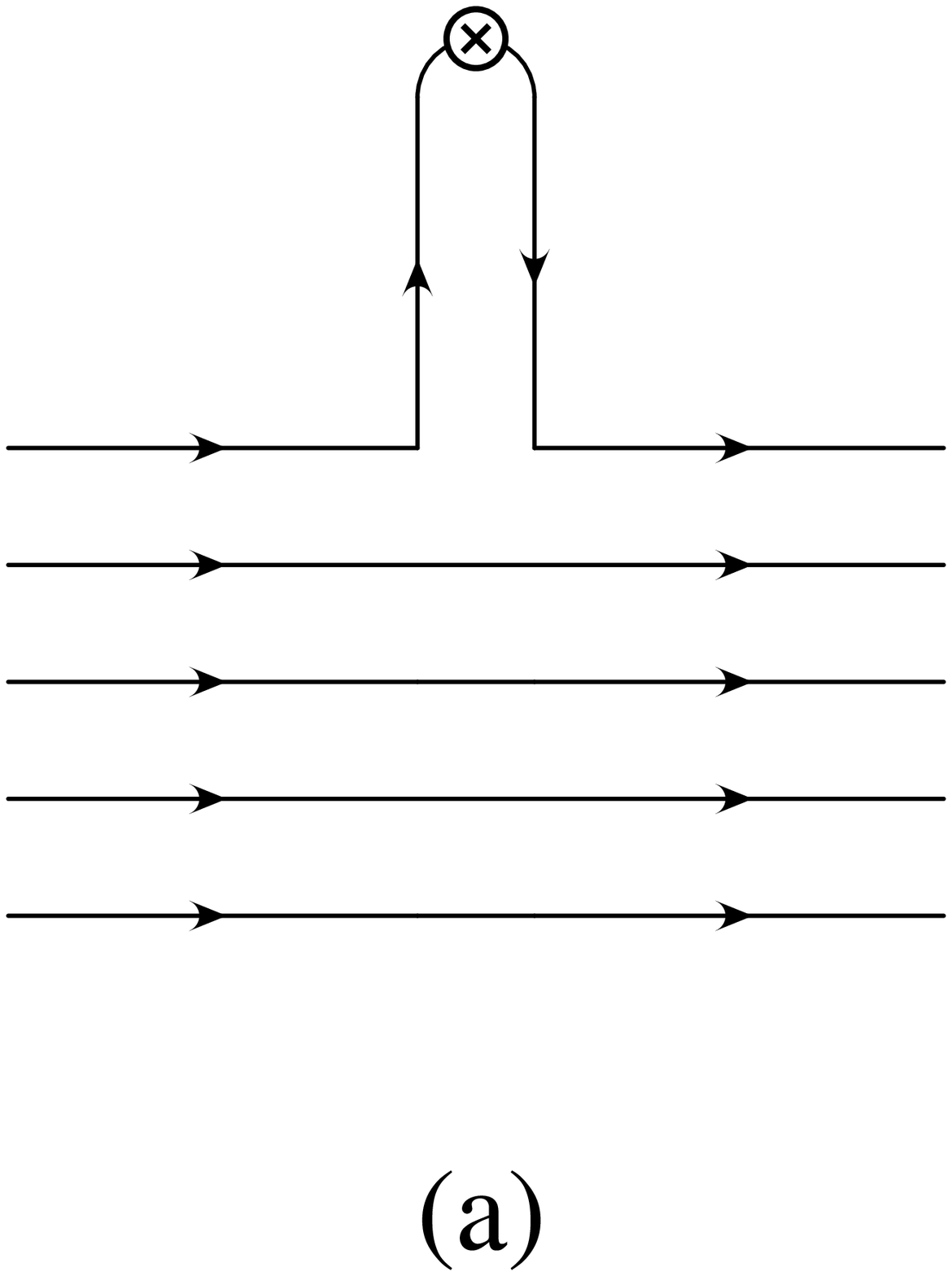}\qquad
\epsfxsize=1.2in\epsffile{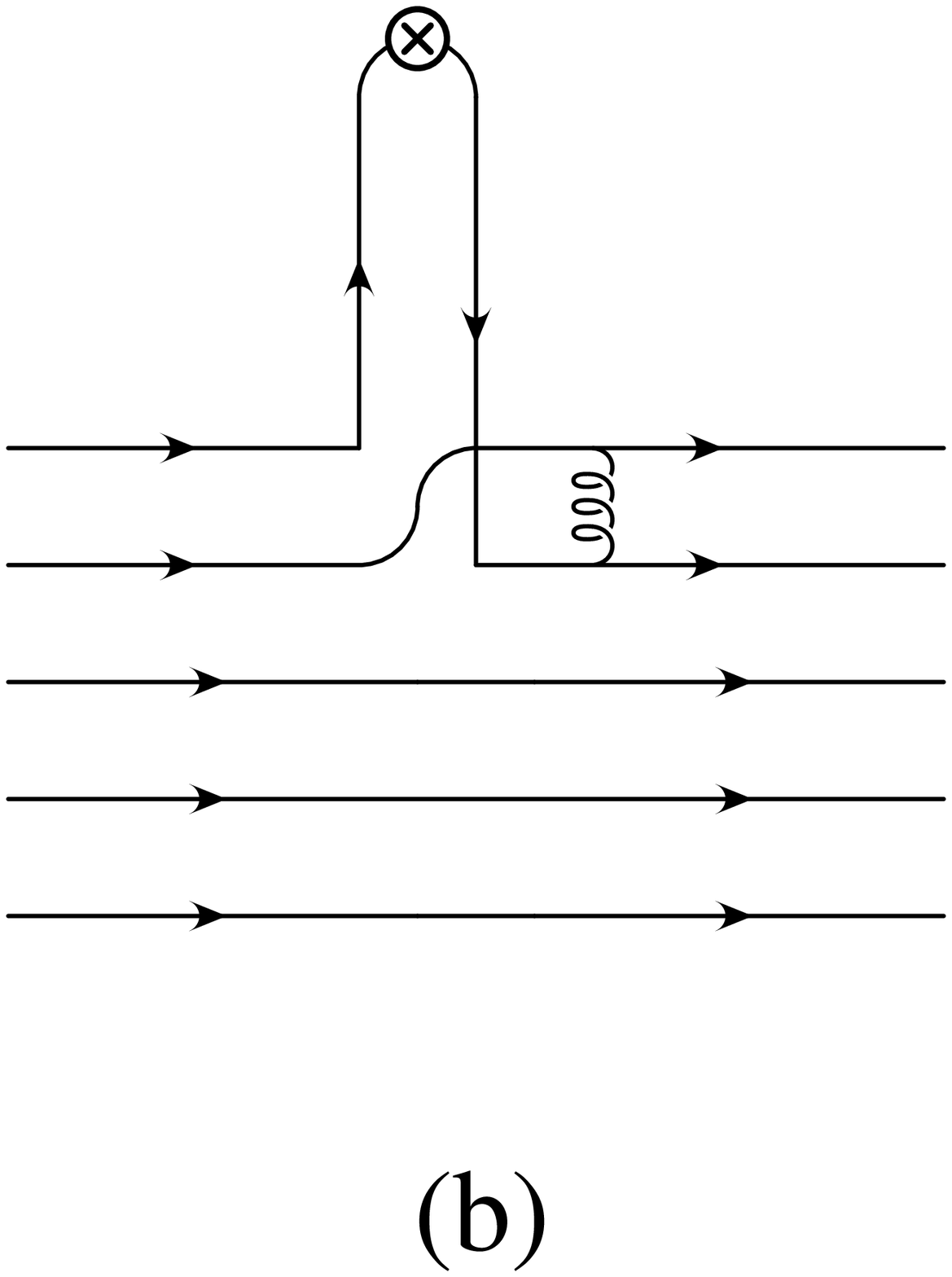}}
\caption{Quark-gluon diagrams for baryon coupling to a meson.}
\end{figure}

\begin{figure}\label{pib}
\centerline{\epsfxsize=2.0in\epsffile{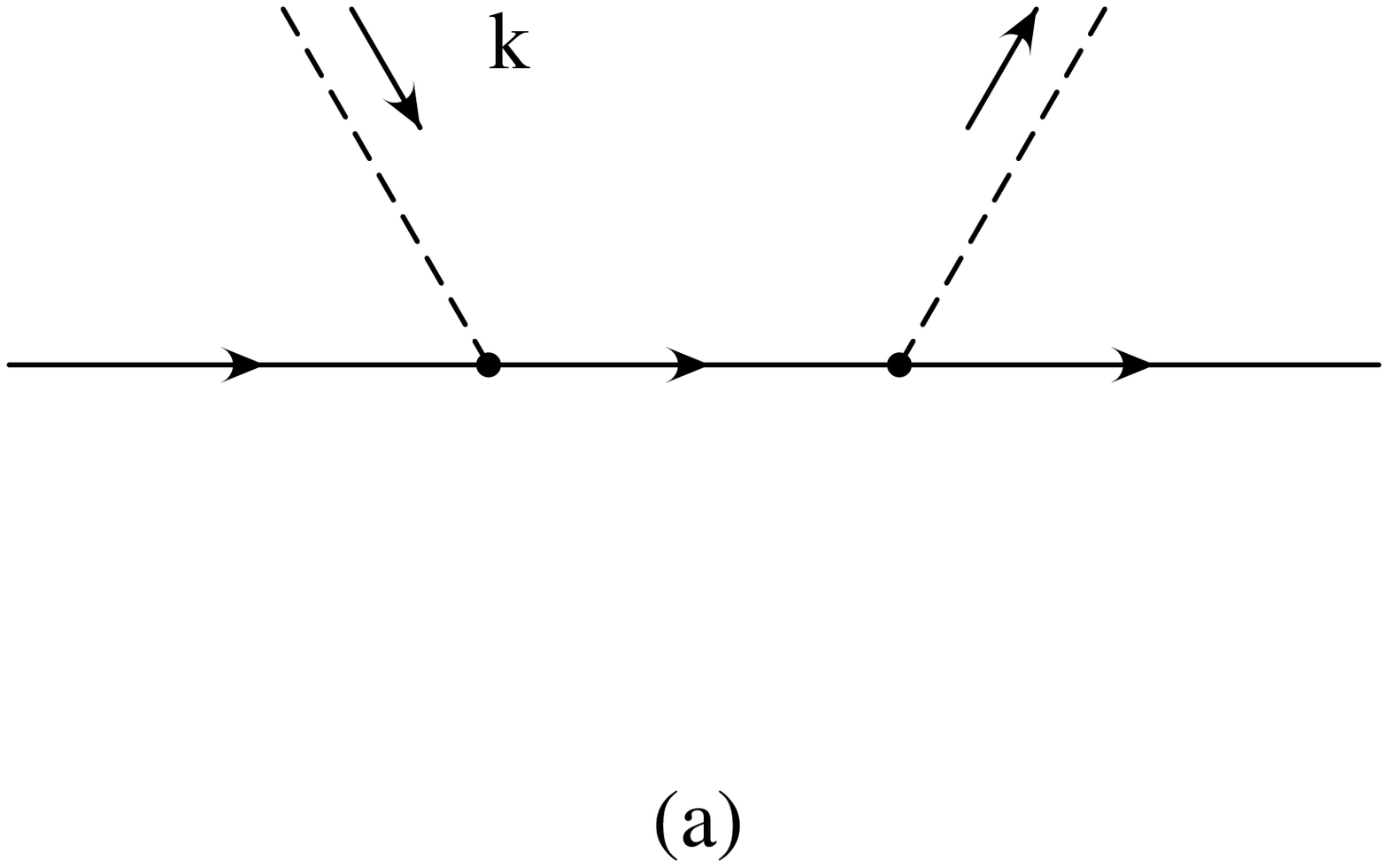}
\qquad\epsfxsize=2.0in\epsffile{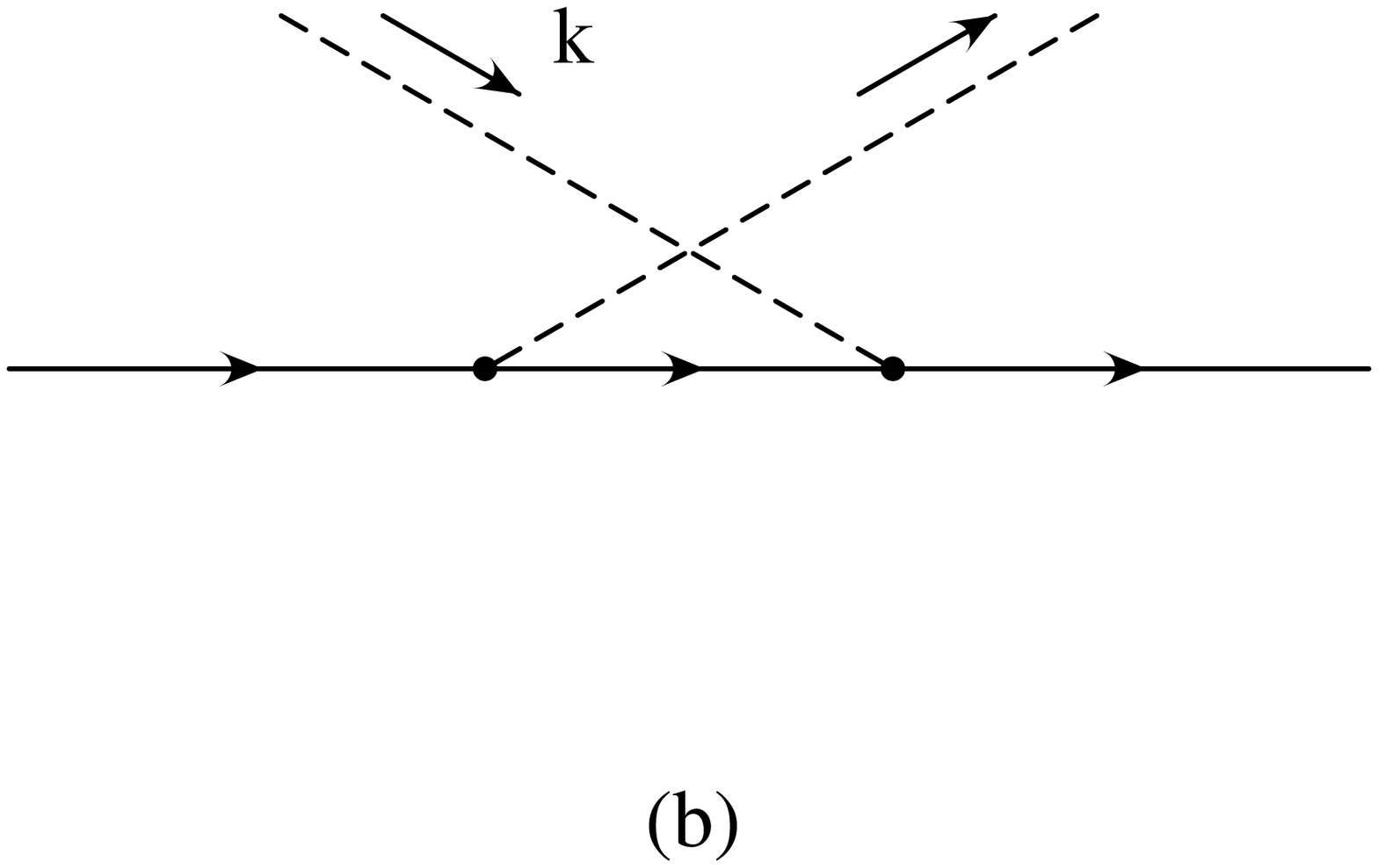}}
\caption{Leading order diagrams for the scattering $B + \pi \rightarrow 
B^\prime + \pi$.}
\end{figure}

\begin{figure}\label{youngtableau}
\centerline{$$\nbox$$}
\caption{$SU(2N_F)$ spin-flavor representation for the ground state baryons.
The Young tableau has $N_c$ boxes.}
\end{figure}

\begin{figure}
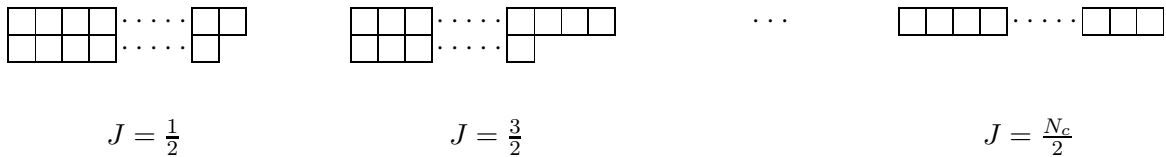
\label{baryontower}
\centerline{ $\nboxF$ \hskip.5in $\nboxE$ \hskip.5in
$\ndots$ \hskip.5in $\nbox$ }
\vskip.25in
\centerline{ ${J= \frac 1 2}$ \hskip1.35in 
${J= \frac 3 2}$ \hskip2.35in
${J= \frac {N_c} 2}$}
\caption{$SU(2) \otimes SU(N_F)$ representations for the ground state
baryons.  The $SU(2N_F)$ representation decomposes into a 
tower of baryon states with 
with $J=\frac 1 2$, $\frac 3 2$, $\ldots$, $\frac {N_c} 2$.  Each Young
tableau has $N_c$ boxes.}
\end{figure}

\setlength{\unitlength}{6mm}

\begin{figure}\label{hyperfine}
\centerline{\hbox{
\begin{picture}(10,7.725)(-1.9,-0.525)
\def\level{\line(1,0){5}}
\thicklines
\put(0,6.4){\level}
\put(0,4.9){\level}
\put(0,3.6){\level}
\put(0,2.5){\level}
\put(0,1.6){\level}
\put(0,0.9){\level}
\put(0,0.4){\level}
\put(0,0.1){\level}
\put(0,0){\level}
\thinlines
\put(5.5,6.4){\line(1,0){1}}
\put(5.5,4.9){\line(1,0){1}}
\put(5.5,0.125){\line(1,0){1}}
\put(5.5,-0.025){\line(1,0){1}}
\put(-1.9,0){\line(1,0){1}}
\put(-1.9,6.4){\line(1,0){1}}
\put(6,5.65){\makebox(0,0){$1$}}
\put(7.5,0){\makebox(0,0){$1/N_c$}}
\put(-1.65,3.05){$N_c$}
\put(-1.4,3.6){\vector(0,1){2.7}}
\put(-1.4,2.8){\vector(0,-1){2.7}}
\put(6,7.2){\vector(0,-1){0.8}}
\put(6,4.1){\vector(0,1){0.8}}
\put(6,-0.525){\vector(0,1){0.5}}
\put(6,0.625){\vector(0,-1){0.5}}
\end{picture}
}}
\caption{Hyperfine mass splittings for the tower of large-$N_c$ baryon
states with $J=\frac 1 2, \frac 3 2, \cdots, \frac {N_c} 2$.  The $J^2/N_c$
operator leads to a mass splitting of $O(1/N_c)$ 
between baryons with spins $J \sim O(1)$,
and to a mass splitting of $O(1)$ between baryons with spins $J \sim O(N_c)$.
The mass splitting between the baryon states with $J= \frac {1} 2$ 
and $J= \frac {N_c} 2$ is $O(N_c)$.}
\end{figure}

\begin{figure}\label{bpbpp}
\centerline{\epsfxsize=2.0in\epsffile{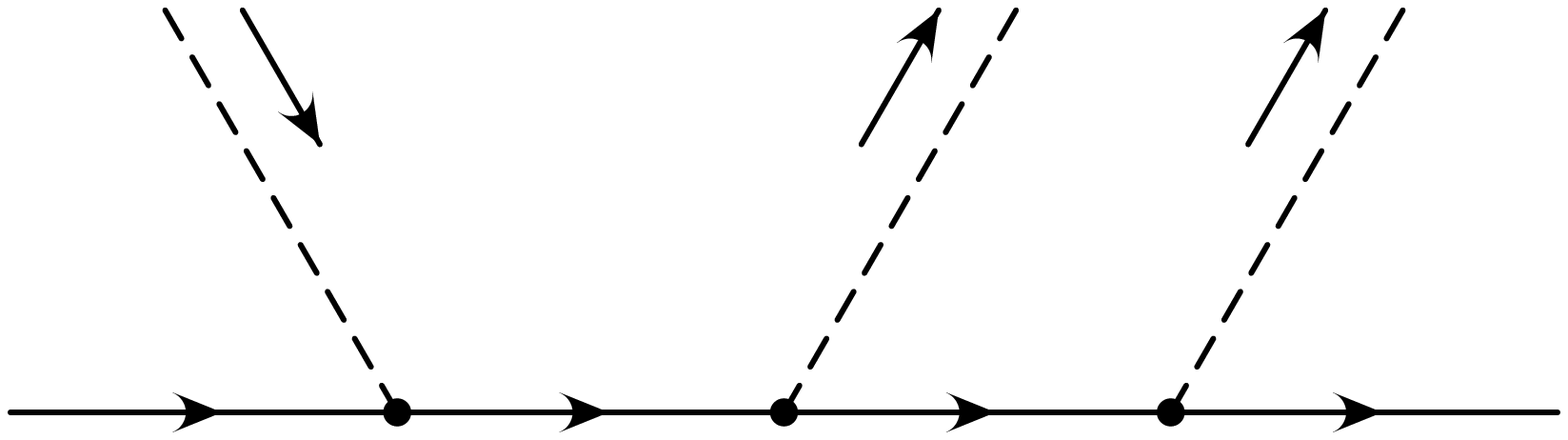}}
\caption{A leading order diagram for the scattering $B+ \pi \rightarrow 
B^\prime + \pi + \pi$.  Additional tree graphs with all permutations of the
pion vertices contribute at leading order.}
\end{figure}

\def\onedot{\makebox(0,0){$\scriptstyle 1$}}
\def\twodot{\makebox(0,0){$\scriptstyle 2$}}
\def\threedot{\makebox(0,0){$\scriptstyle 3$}}
\def\fourdot{\makebox(0,0){$\scriptstyle 4$}}

\setlength{\unitlength}{3mm}

\begin{figure}\label{spin12}
\centerline{\hbox{
\begin{picture}(20.79,18)(-10.395,-8)
\multiput(-1.155,10)(2.31,0){2}{\onedot}
\multiput(-2.31,8)(4.62,0){2}{\onedot}
\multiput(-3.465,6)(6.93,0){2}{\onedot}
\multiput(-4.62,4)(9.24,0){2}{\onedot}
\multiput(-5.775,2)(11.55,0){2}{\onedot}
\multiput(-6.93,0)(13.86,0){2}{\onedot}
\multiput(-8.085,-2)(16.17,0){2}{\onedot}
\multiput(-9.24,-4)(18.48,0){2}{\onedot}
\multiput(-10.395,-6)(20.79,0){2}{\onedot}
\multiput(-9.24,-8)(2.31,0){9}{\onedot}
\multiput(0,8)(2.31,0){1}{\twodot}
\multiput(-1.155,6)(2.31,0){2}{\twodot}
\multiput(-2.31,4)(2.31,0){3}{\twodot}
\multiput(-3.465,2)(2.31,0){4}{\twodot}
\multiput(-4.62,0)(2.31,0){5}{\twodot}
\multiput(-5.775,-2)(2.31,0){6}{\twodot}
\multiput(-6.93,-4)(2.31,0){7}{\twodot}
\multiput(-8.085,-6)(2.31,0){8}{\twodot}
\end{picture}
}}
\caption{$SU(3)$ flavor weight diagram for
the spin-${1 \over 2}$ baryons.  The numbers
denote the multiplicity of the weights.  The long side of the weight diagram
contains ${1 \over 2}\left( N_c + 1 \right)$ weights.}
\end{figure}

\begin{figure}\label{spin32}
\centerline{\hbox{
\begin{picture}(20.79,18)(-8.085,-8)
\multiput(-1.155,10)(2.31,0){4}{\onedot}
\multiput(-2.31,8)(9.24,0){2}{\onedot}
\multiput(-3.465,6)(11.55,0){2}{\onedot}
\multiput(-4.62,4)(13.86,0){2}{\onedot}
\multiput(-5.775,2)(16.17,0){2}{\onedot}
\multiput(-6.93,0)(18.48,0){2}{\onedot}
\multiput(-8.085,-2)(20.79,0){2}{\onedot}
\multiput(-6.93,-4)(18.48,0){2}{\onedot}
\multiput(-5.775,-6)(16.17,0){2}{\onedot}
\multiput(-4.62,-8)(2.31,0){7}{\onedot}
\multiput(0,8)(2.31,0){3}{\twodot}
\multiput(-1.155,6)(6.93,0){2}{\twodot}
\multiput(-2.31,4)(9.24,0){2}{\twodot}
\multiput(-3.465,2)(11.55,0){2}{\twodot}
\multiput(-4.62,0)(13.86,0){2}{\twodot}
\multiput(-5.775,-2)(16.17,0){2}{\twodot}
\multiput(-4.62,-4)(13.86,0){2}{\twodot}
\multiput(-3.465,-6)(2.31,0){6}{\twodot}
\multiput(1.155,6)(2.31,0){2}{\threedot}
\multiput(0,4)(4.62,0){2}{\threedot}
\multiput(-1.155,2)(6.93,0){2}{\threedot}
\multiput(-2.31,0)(9.24,0){2}{\threedot}
\multiput(-3.465,-2)(11.55,0){2}{\threedot}
\multiput(-2.31,-4)(2.31,0){5}{\threedot}
\multiput(2.31,4)(2.31,0){1}{\fourdot}
\multiput(1.155,2)(2.31,0){2}{\fourdot}
\multiput(0,0)(2.31,0){3}{\fourdot}
\multiput(-1.155,-2)(2.31,0){4}{\fourdot}
\end{picture}
}}
\caption{$SU(3)$ flavor weight diagram for
the spin-${3 \over 2}$ baryons. The numbers
denote the multiplicity of the weights.  The long side of the weight diagram
contains ${1 \over 2}\left( N_c - 1 \right)$ weights.}
\end{figure}
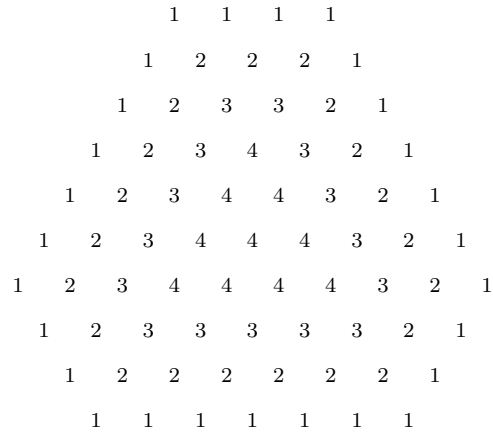

\begin{figure}
\moveright1cm\hbox{\epsfxsize=10 cm \epsffile{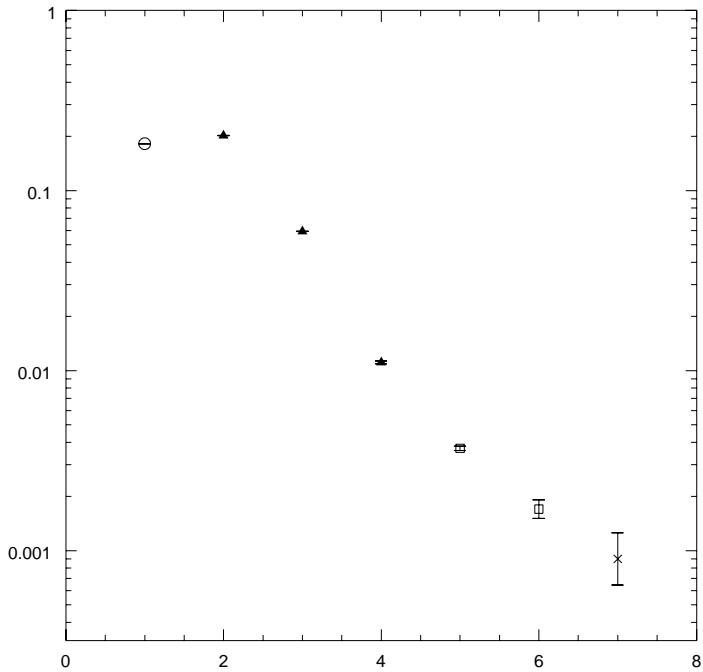}}
\caption{Isospin-averaged baryon mass combinations of Jenkins $\&$
Lebed~\cite{jl}. The error bars are experimental.  
A hierarchy of baryon masses in $1/N_c$ and $SU(3)$ flavor
symmetry breaking $\epsilon \sim 0.3$ is predicted by the theoretical analysis. 
The open circle is a mass combination of order $1/N_c^2$; the
three solid triangles are mass relations of order $\epsilon/N_c$,
$\epsilon/N_c^2$, and $\epsilon/N_c^3$; the open squares
are relations of order $\epsilon^2/N_c^2$,
and $\epsilon^2/N_c^3$; and the $\times$ is a relation of
order $\epsilon^3/N_c^3$.}
\end{figure}

\begin{figure}
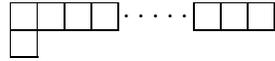
\label{excited}
\centerline{$$\nboxA$$}
\caption{$SU(2N_F)$ spin-flavor representation for the first excited baryons.
The Young tableau has $N_c$ boxes.}
\end{figure}

\end{document}